# Frustration-Mediated Crossover from Long-Range to Short-Range Magnetic Ordering in $Y_{1-x}Lu_xBaCo_4O_7$

Sevda Sahinbay[1,2,3], Hong Zheng[2], John F. Mitchell[2], Stephan Rosenkranz[2], Omar Chmaissem[2,3]

[1] *Department of Engineering Physics, Istanbul Medeniyet University, Istanbul 34700, Türkiye*

[2] *Materials Science Division, Argonne National Laboratory, Lemont, IL 60349, USA*

[3] *Department of Physics, Northern Illinois University, DeKalb, IL 60115, USA*

**Abstract**

We present a comprehensive magnetic and structural phase diagram for $Y_{1-x}Lu_xBaCo_4O_7$, established through neutron diffraction and magnetization measurements. Our results outline the evolution of various nuclear structures, encompassing both orthorhombic and monoclinic symmetries, in response to changes in composition and temperature. The phase transition temperature of the orthorhombic phase, $T_{s1}$, decreases with increasing Lu content from 310 K for x = 0.0 to 110 K for x = 1.0. In Lu-rich compositions (0.7 ≤ x ≤ 1.0), first-order structural transitions are observed with coexisting and competing orthorhombic P$bn$2$_1$ and metastable monoclinic C$c$ phases. Composition- and temperature-dependent refinements of the magnetic structure reveal an antiferromagnetic arrangement of Co spin pairs, below the magnetic transition temperatures of the Y-rich compositions, in the *ab* plane within both the triangular and Kagomé layers. A gradual suppression of long-range magnetic order is observed with increasing the Lu content, accompanied by the development of short-range magnetic correlations present in all the samples.

## I. Introduction

Geometrically frustrated systems represent a unique class of materials in which the configuration of magnetic moments on the crystal lattice leads to frustration. This frustration arises from an inherent mismatch between magnetic interactions and lattice geometry, preventing the establishment of a unique ground state due to the impossibility of satisfying all interactions simultaneously. Various lattice geometries such as triangular, Kagomé, and pyrochlore lattices [1–11], are associated with geometric frustration. A characteristic feature of these materials is the persisting absence of a conventional ordered magnetic state, even at the lowest temperatures. Instead, observations often reveal the presence of short-range or spin-liquid-like correlations.

The family of $R$BaCo$_4$O$_7$ cobaltites, denoted as $R$-114 [with $R$ = Ca, In, Y, or a rare earth ($RE$) element] serves as an exemplary system for investigating competing short-range (SRO) and long-range (LRO) magnetic orders resulting from geometric frustration within a trigonal structural framework. The trigonal structure is best described as a stack of alternating two-dimensional Kagomé and triangular layers, with cobalt ions occupying the center of corner-shared oxygen tetrahedra, as illustrated in Fig. 1. Charge balance necessitates the presence of a mixed Co oxidation state, with Co$^{2+}$ and Co$^{3+}$ ions randomly distributed among the tetrahedral centers of two independent sites. For compositions with a trivalent $R$ ion (i.e., $R$ = In, Y, or a rare earth ($RE$) element), an average Co oxidation state of 2.25+ is expected, along with a three-to-one Co$^{2+}$/Co$^{3+}$ ratio. In the case of Ca-114 (where $R$ is a divalent ion), an equal number of Co$^{2+}$ and Co$^{3+}$ ions is present in a 1:1 ratio, owing to the average Co oxidation state increasing to 2.5+ [12–15].

$R$-114 materials undergo a first-order trigonal P$31c$ to orthorhombic P$bn2_1$ structural phase transition at a temperature, $T_{S1}$, slightly above room temperature. The mechanism responsible for breaking the three-fold rotational symmetry of the trigonal structure was linked to the temperature-dependent behavior of severely under-bonded Ba ions [16], or potentially to charge redistribution of the Co ions [17,18]. Irrespective of the origin of the phase transition, $T_{S1}$ displays a linear dependence on the ionic radius of the $R^{3+}$ cation [19].

On the other hand, the *R*-114 system exhibits various magnetic properties at temperatures below 100 K, strongly influenced by the choice of the *R*-cation. Magnetization and magnetic relaxation measurements reveal spin glass-like freezing behavior in InBaCo$_4$O$_7$ [20], for example, while clear ferrimagnetic order below 70 K is demonstrated in CaBaCo$_4$O$_7$ [21,22]. YBaCo$_4$O$_7$ exhibits fast decaying spin-spin correlations indicative of spin liquid properties [23], coupled with long-range antiferromagnetic (AFM) ordering below 106 K ($T_N$). Spin reorientation below 50 K is suggested by neutron diffraction [13,24].

Combining neutron powder and single crystal diffraction, Chapon *et al.* [13] and Khalyavin *et al.* [24] proposed two slightly modified magnetic ordering models for YBaCo$_4$O$_7$ using the Shubnikov magnetic space group P2$_1$', with the main difference being an out-of-plane component allowed for the Co spins in the Kagomé layers [24]. Conversely, Hoch *et al.* [25], using $^{59}$Co NMR, suggested an ordering in which the Co magnetic moments in the triangular layer are similar to the neutron diffraction model, while those in the Kagomé layer all lie within the *ab* plane and form a different pattern. In addition to the long-range magnetic order observed in Y-114 single crystals, the presence of significant short-range magnetic correlations below 250 K [13] was suggested, with these correlations persisting down to 2 K [24].

In prior research, we documented the observation of two successive first-order phase transitions in Lu-114, depending on the cooling speed to base temperatures. Fast cooling suppresses the low temperature orthorhombic P*bn*2$_1$ structure, allowing the transition from the trigonal P31*c* symmetry to proceed directly to a metastable monoclinic C*c* phase below 160 K, which remains stable down to 2 K. Slow cooling protocols proved inadequate in removing the rejected heat swiftly enough to suppress the orthorhombic phase, resulting in coexisting orthorhombic and monoclinic phases with varying phase fractions dependent on the cooling rate [12]. To the best of our knowledge, the intermediate monoclinic phase is unique to Lu-114 and has not been observed in other *R*-114 systems. Additionally, no long-range magnetic ordering is observed in Lu-114 [12,14], except for broad diffuse magnetic scattering intensities appearing after an extended waiting time at low temperatures, indicating short-range magnetic correlations below ~95 K [12].

The distinct structural and magnetic features of the Y and Lu members of the *R*-114 family have motivated us to explore the entire $Y_{1-x}Lu_xBaCo_4O_7$ solid solution system. Our goal is to understand how lattice and spin order evolve across the series. In this study, we present a comprehensive phase diagram for $Y_{1-x}Lu_xBaCo_4O_7$, revealing a crossover from long-range magnetic ordering to short-range magnetic correlations as the Lu content increases. The temperature of the trigonal to orthorhombic structural transition, $T_{s1}$, progressively decreases with the rising Lu concentration. The metastable C*c* phase is exclusively observed in compositions where $0.6 < x \leq 1.0$. On the Y-rich side of the phase diagram, Rietveld refinements reveal antiferromagnetic coupling of four cobalt spin pairs in both the Kagomé and triangular layers, with their directions confined to the *ab* plane. Within each Kagomé triangle, two spins exhibit ferromagnetic coupling, while the other are coupled antiferromagnetically, forming what resembles "stripes" along one of the unit cell's *ab* plane diagonals. Magnetic short-range correlations are observed in all samples with their magnitude gradually increasing as the Lu content rises.

## II. Experimental Details

Solid solutions of $Y_{1-x}Lu_xBaCo_4O_7$ with Δx intervals of 0.1 were synthesized using conventional solid-state methods. To eliminate moisture from $Y_2O_3$ and $BaCO_3$ precursors, the powders were dried separately, with $Y_2O_3$ dried overnight at 1000 °C in flowing oxygen and $BaCO_3$ in a box oven at 180 °C. High purity powders of $Lu_2O_3$ (99.999%), $Y_2O_3$ (99.995%), $Co_3O_4$ (99.99%) and $BaCO_3$ (99.99%) were combined in stoichiometric ratios. The mixtures underwent sintering in air at 1000 °C, 1050 °C, and 1150 °C in successive 24 h cycles, with intermediate grindings. At the conclusion of the final sintering cycle, the samples were rapidly quenched to room temperature and subsequently annealed at 600 °C for 12 h under argon, resulting in the formation of the final stoichiometric $Y_{1-x}Lu_xBaCo_4O_{7.0}$ phases.

Magnetic susceptibility measurements were conducted on powder samples enclosed in small gelatin capsules utilizing a Quantum Design PPMS magnetometer (Quantum Design, San Diego, CA, USA). Samples were measured using zero-field-cooled on warming (ZFC-W), field-cooled on cooling (FC-C), and field-cooled on warming (FC-W) protocols, all conducted under a magnetic field of 1 T. DC magnetization measurements were performed during both the warming and cooling processes at a rate of 2 K/min.

High-resolution neutron powder diffraction data were collected using the time-of-flight (TOF) diffractometer POWGEN [26] at the Spallation Neutron Source, Oak Ridge National Laboratory. Data for Lu-substituted samples were collected in two frames with center wavelengths of 1.066 Å and 3.731 Å, while data for $YBaCo_4O_7$ were collected in a single frame with a center wavelength of 1.5 Å. Neutron diffraction experiments were performed on warming.

To account for the metastable C*c* phase fraction's dependence on the cooling rate, with the highest fraction achieved by fast cooling, samples were cooled as rapidly as possible to base temperature using POWGEN's automatic sample changer. This process typically took about 80 minutes to reach 12 K, equivalent to approximately 3.5 K/min. For $LuBaCo_4O_7$, this cooling rate was previously demonstrated to be sufficient for achieving 100% phase conversion to the metastable monoclinic phase [12].

Nuclear and magnetic structures were determined using Rietveld refinement methods as implemented in the General Structure Analysis System (GSAS II) [27,28].

## III. Results and Discussion

### A. Phase Diagram

We mapped out the structural and magnetic phase diagram for the entire $Y_{1-x}Lu_xBaCo_4O_7$ series as shown in Fig. 2. Consistent with earlier findings, the 300 K structure of $YBaCo_4O_7$ is orthorhombic, undergoing a structural transition from trigonal symmetry at a slightly elevated temperature [13]. Substituting Lu for Y, ranging from partial to full substitution, continuously suppresses the transition temperature, $T_{S1}$, down to 110 K (for x = 1). In all Lu-substituted samples, the room temperature structure falls within the trigonal phase regime.

The values of $T_{S1}$, determined through zero-magnetic-field neutron diffraction and Rietveld refinements (depicted as magenta triangles in Fig. 2), align well with those extracted from the first derivatives of the magnetization curves (represented by black squares in Fig. 2), despite the latter measurements being under 1T magnetic fields. The light violet region in the phase diagram delineates the equilibrium-state orthorhombic P*bn*2$_1$ phase. On the Lu-rich side of the phase diagram (0.7 ≤ x ≤ 1.0), two successive structural transitions to monoclinic C*c* and then to orthorhombic P*bn*2$_1$ are observed. For completeness, a small region in cyan is included in the phase diagram, indicating where the metastable monoclinic phase would exist if the sample were cooled under non-equilibrium thermodynamic conditions (*i.e.*, fast cooling rates).

As mentioned earlier, oxygen stoichiometric YBaCo$_4$O$_{7.0}$ exhibits a long-range ordered magnetic structure below 110 K ($T_N$) [13]. A slight increase in oxygen content, as in YBaCo$_4$O$_{7.1}$ for example, completely suppresses both the nuclear and magnetic transitions with the structure remaining trigonal down to the lowest measured temperature at 6 K [29]. Sharp and intense magnetic peaks shown in Fig. 3 and the supplementary Figure 1 in the supplementary material (SM) [30] provide strong evidence for the full and exact oxygen stoichiometry (*i.e.*, 7.0 oxygen atoms per formula unit) in our Y-rich compositions.

As shown in Fig. 2, Lu substitution progressively weakens the magnetic order and reduces the onset of the magnetic transition temperature to 70 K for compositions with x up to 0.5. Samples with higher Lu contents (x > 0.5) exhibit no long-range magnetic order, which is suppressed in favor of short-range magnetic correlations, as seen in the neutron diffraction data displayed in Fig. 3(b), for example. Short-range magnetic correlations manifest as diffuse scattering intensities at ~120 K and below for the x = 0.0 sample (supplementary Figure 1 [30]) and below ~100 K for all Lu substituted samples, including those displaying long-range magnetic order.

Due to relatively coarse substitution increments (Δx = 0.1) in our solid solution materials, the exact composition between x = 0.5 and 0.6 at which the transition from long-range to short range correlations occurs, both in terms of temperature and Lu content, cannot be precisely determined. Nevertheless, reasonable magnetic structural refinements at 12 K were performed for the x ≤ 0.5 samples, despite the gradual and fast suppression (Fig. 3) and broadening of the magnetic peaks as Lu content increased (see inset of Fig. 2).

On the other hand, the onset of short-range magnetic order (SRO), occurring simultaneously with long-range magnetic order, was determined by fitting temperature-dependent integrated intensities of several magnetic reflections as shown in Fig. 4, for example. The Boltzmann function (solid red curves) was used for fitting, and the second derivatives of the fits (insets in Fig. 4) were used to determine the inflection points corresponding to the onset of SRO in our bulk polycrystalline materials. Further insights into short-range magnetic correlations, beyond the scope of this paper, would necessitate inelastic and quasi-elastic neutron measurements from single crystals.

**B. Structural Properties**

Our phase diagram comprehensively delineates most of the known nuclear phases in the *R*-114 system, encompassing the metastable monoclinic C*c* phase on the Lu-rich side. However, we observed the absence of another potential low temperature monoclinic phase (P$2_1$) on the Y-side, reported for YBaCo$_4$O$_7$ [13,24] based on peak splitting of the nuclear (221) reflection using high resolution synchrotron x-ray diffraction.

In our neutron diffraction data for all samples, no evidence of peak splitting or broadening indicative of a monoclinic phase on the Y-rich side was observed. Despite the absence of apparent monoclinic distortions, attempts to perform meaningful nuclear structure Rietveld refinements using the P$2_1$ symmetry were not possible due to the low symmetry requiring 25 independent atoms and 3 degrees of freedom each. Therefore, while we cannot conclusively rule out the possible existence of subtle monoclinic distortions at a scale smaller than the resolution limit of POWGEN, our nuclear structural refinements were successfully conducted using the orthorhombic P*bn*$2_1$ symmetry between T$_{S1}$ and 12 K.

The first-order P*bn*$2_1$ structural phase transition observed at both ends of the phase diagram extends over a temperature range toward the center of the diagram. In the vicinity of the onset temperature of the phase transition, T$_{S1}$, polycrystalline samples with $0.1 \leq x \leq 0.7$ exhibit biphasic behavior. Two-phase refinements were conducted to characterize this coexistence, as reflected in the relatively large uncertainties represented by error bars over magenta triangles on the phase diagram. A contour map displayed in Fig. 5 illustrates this phase coexistence during the phase transition, with additional plots for other samples displayed in the supplementary Figure 2 [30].

Similar to LuBaCo$_4$O$_7$ [12], consecutive low-temperature regions exhibiting biphasic and single-phase monoclinic C$c$ structures are observed by neutron diffraction for samples with x ≥ 0.7 when measured on warming. While a single monoclinic C$c$ phase is achieved in the x = 1.0 sample (LuBaCo$_4$O$_7$) at 12 K, employing the same fast-cooling protocol at the same instrument (POWGEN) for x = 0.7, 0.8 and 0.9 resulted in variable biphasic mixtures at 12 K. An example contour map for the x = 0.9 sample in Fig. 6(a) illustrates the temperature-dependent evolution of mixed C$c$ and P$bn2_1$ diffraction peaks at 1.565 Å (indexed as monoclinic 400) and 1.57 Å (indexed as orthorhombic 145 and 253), respectively. The intense middle peak represents overlapped reflections from both phases, merging into a single trigonal P31$c$ (220) reflection at temperatures above T$_{S1}$. No trace of the monoclinic C$c$ phase was detected within the resolution limit of POWGEN in any of the remaining samples with x ≤ 0.6. Additional neutron diffraction plots for x = 0.7 and 0.8 are shown in the supplementary Figure 3 [30], along with the refined phase fractions by weight of the identified phases.

The weight fractions of the C$c$ and P$bn2_1$ mixed phases for x = 0.9 remain relatively unchanged between 12 K and ~50 K (Region I in Fig. 6). The P$bn2_1$ phase gradually becomes dominant from ~50 K to ~100 K (within Region II) due to the thermally-induced (kinetic) conversion upon warming of the metastable C$c$ phase to the thermodynamically stable P$bn2_1$ phase. This behavior mirrors that of pure Lu-114 (x = 1.0), although in the latter composition, a 100% monoclinic C$c$ phase was achieved at 10 K under similar conditions [12].

Upon further heating above ~100 K, the phase fractions reverse trends as the system approaches the stability conditions of the C$c$ phase near 120 K. The C$c$ phase then remains present in a single-phase form until its suppression at ~160 K, with the structure transitioning to the trigonal P31$c$ symmetry. The inversely related C$c$ and P$bn2_1$ phase fractions change continuously with increasing Y content, as shown in the inset of Fig. 6(b). Structural models displaying the subtleties of the competing C$c$ and P$bn2_1$ phases are illustrated in the supplementary Figure 4 [30].

The unit cell lattice parameters and unit cell volume at 12 K are shown in Fig. 7 as a function of Lu substitution. The unit cell volume, as well as the in-plane *a* and *b* lattice parameters, display the anticipated linear behavior according to Vegard's law's [31]. This linear behavior arises from site-sharing ions with contrasting ionic radii, such as the six-coordinated Y and Lu ions, where $r_Y$ = 0.90 Å and $r_{Lu}$ = 0.861 Å [32]. It is worth noting that the full replacement of Y by the smaller Lu ion shrinks the unit cell volume by about 1.7%, thus, increasing the magnetic interaction frustrations.

The evolution of the unit cell volume and unit cell lattice parameters as a function of temperature is illustrated in Fig. 8 and the supplementary Figure 5 [30], respectively, for all three nuclear phases observed in our samples. While the Y-rich compositions (x = 0.1 and 0.2) display typical thermal expansion behavior, a clear inflection point at ~120 K is noticeable, starting with x = 0.5 for the P*bn*$2_1$ unit cell volume (depicted by black triangles). Below this temperature, the volume becomes nearly flat and remains unchanged. The response of the short *a*-axis to temperature is more pronounced, exhibiting negative thermal expansion behavior for the x = 0.1 and x = 0.2 samples. No changes are observed along the *b*-axis direction, and the self-compensated combination of all three parameters lead to the unit cell volume behavior shown in Fig. 8.

C. **Magnetic properties**

The transition between trigonal and orthorhombic symmetries, while structurally driven, manifests as a kink in the magnetization data collected under a magnetic field of 1 T (see Fig. 9). This anomalous behavior, explained in previous work [16], is attributed to modified exchange interactions among the Co ions in response to a discontinuous rearrangement of the structure. Consistent with the gradual suppression of $T_{s1}$ observed in our neutron diffraction results, the kink shifts to lower temperatures with increasing Lu content.

The long-range AFM ordering for Y-rich compositions is subtly indicated by a change in slope at ~100 K in the magnetization data (Fig. 9), corresponding to the temperature where the magnetic peaks appear in the neutron data. A pronounced spin reorientation [13,24] hump observed at ~50 K in YBaCo$_4$O$_7$ broadens and shift to lower temperatures with increasing Lu content before sharpening for samples with x ≥ 0.7, as observed in the ZFC-W magnetization data, Fig. 9. This sharp peak, unrelated to spin reorientation in the absence of long-range magnetic ordering, correlates well with the emergence of the metastable C$c$ phase and can be significantly suppressed with very slow cooling rates. Although cooling-rate-dependent measurements similar to those reported in [12] were not performed for our substituted samples, we note the distinct character of the peak – being sharp for slowly cooled Lu-rich samples and broad for Y-rich samples where long-range magnetic ordering and the P$bn2_1$ phase dominate at low temperatures, see inset of Fig. 6(b).

**D. Magnetic Structures**

In the Y-rich samples, strong long-range magnetic peaks are observed at d-spacings ~5.36 Å, 4.61 Å, 3.96 Å, 3.83 Å, 3.64 Å, and 3.21 Å (see Fig. 3) consistent with those reported in Ref. [24]. The continuous enhancement of short-range magnetic correlations, at the expense of suppressed long-range order, is reflected in the weakened peak intensities with increasing Lu-content. Two additional magnetic peaks at 6.3 Å (100 reflection) and 7.4 Å (011 reflection) are specific to the x = 0.0 sample. Even a minimal 10% Lu substitution at the Y site is enough to suppress these two peaks, as shown in the supplementary Figure 6 [30]. Thus, the magnetic structure of the Y-only sample may differ from that of the Lu-substituted ones, until only broad magnetic diffuse scattering intensities are observed around 4.6 Å for x > 0.5, Fig. 3.

Figure 10 presents a limited range of neutron diffraction data for x = 0.1, highlighting several magnetic peaks below 100 K. Additional neutron diffraction data for x = 0.2, 0.4, 0.5 and 0.6 are shown in the supplementary Figures 7 and 8 [30]. Group theory calculations using P$bn2_1$ and the propagation vector k = (0,0,0) on the Bilbao Crystallographic server integrated with GSAS II resulted in 11 possible magnetic space groups with the Shubnikov symmetry notations P$b'n'2_1$, P$b'n2_1'$, P$bn'2_1'$, P$bn2_1$, P$c'$, P$n'$, P$c$, P$n$, P$2_1'$, P$2_1$, and P1 in which the primes correspond to time reversal symmetry operations. The magnetic peaks can all be indexed using the first four high-symmetry orthorhombic magnetic space groups; however, their calculated intensities were inadequate and leading to poor fits. All the remaining monoclinic space groups, except P$2_1'$, were equally unsatisfactory. It is worth noting that P$2_1'$ is the same space group suggested in the literature [13,24]. In our final refinements, the orthorhombic P*bn*$2_1$ and monoclinic P$2_1$' space groups were used for refinements of the nuclear and magnetic structures, respectively. Magnetic refinements were initiated with no constraints on the magnetic moments of the eight cobalt atoms defined by P$2_1$'.

In the initial refinement process, the symmetry-allowed Cartesian coordinate components ($M_x$, $M_y$, $M_z$) of the magnetic moments were freely refined, starting from unbiased values deemed reasonable. However, negligible $M_z$ components were observed for all samples and temperatures during stable refinement cycles. Consequently, these out-of-plane components were fixed at zero during the final refinement cycles. Additionally, the refined magnetic moments of specific cobalt pairs (Co1 and Co2 in the triangular layer, and Co3 and Co4, Co5 and Co6, Co7 and Co8 in the Kagomé layer) displayed a strong tendency to couple antiferromagnetically. To reflect this tendency, constraints were applied accordingly.

Table 1 provides the Co atomic positions for x = 0.1 at 12 K and the final magnetic structures at 10-12 K are displayed in Fig. 11. Notably, our model diverges from the published ones [24] in two key aspects: first, the magnetic moment magnitudes of the Kagomé layer's cobalt ions are not constrained to be equal, and second, various Co pairs tend to align antiferromagnetically, particularly the Co7-Co8 (cyan spheres) and Co3-Co4 (green spheres) pairs, giving rise to a distinctive "stripes" appearance along the $[1\bar{1}0]$ direction. An example of best-fit Rietveld refinement plots is displayed in the supplementary Figure 9 [30].

Magnetic refinements for the x = 0.0 sample resulted in two slightly different models (A and B) at 5 K that are visually indistinguishable from a fit inspection or examination of the similar agreement factors of the refinements. Interestingly, this is the only sample for which the refinements converged to two different solutions. Model A, as illustrated in the supplementary Figures 10(a) and (b) [30], agrees with the structures obtained at 12 K for x = 0.1 and x = 0.2, whereas model B shows the spins of the Co3-Co4 pairs rotated in a direction "symmetrical" to that of model A with everything else remaining roughly similar, see supplementary Figures 10(c) and (d) [30]. However, when these two models were imposed on the Lu-substituted samples, a consistent and "unique" solution was always obtained, as described above, and shown in Figs. 11 and 12. The robustness of our final magnetic models was evidenced by the similar solutions obtained for all the long-range magnetically ordered samples irrespective of the initial magnetic moment values. Attempts using constraints as in the published models [13,24,25] resulted in poor fits (see supplementary Figure 11 [30]), but releasing these constraints produced the same magnetic structures shown in Fig. 11.

Stable Rietveld refinements were achieved for all x ≤ 0.5 samples at 12K. Temperature-dependent refinements were only possible for x = 0.1 and 0.2 samples for which the magnetic peaks are relatively strong at higher temperatures. The temperature evolution of Co magnetic ordering for x = 0.1 and 0.2 in the Kagomé and triangular layers is shown in Fig. 12 and the supplementary Figure 12 [30], respectively. Due to limited beamtime, temperature-dependent neutron data were not collected for the x = 0.3 sample. The x = 0.4 and 0.5 samples exhibit small long-range ordered magnetic moments at 12 K, diminishing rapidly with increasing temperature, hindering meaningful refinements of temperature-dependent magnetic structures. The gradual suppression and broadening of magnetic peak intensities, coupled with coarse substitution increments, make it difficult to delineate a clear boundary separating long-range magnetic order from short-range correlations as a function of Lu content.

The temperature dependence of magnetic structures of x = 0.1 and 0.2 correlates well with the hump observed in their 1T magnetization data. This correlation is evident in the non-monotonic evolution of the magnetic structure and the expected hump-related spin reorientation in both the Kagomé and triangular layers. The magnetic structure for x = 0.1 at 12 K and 20 K and for x = 0.2 at 12 K (well below the magnetization hump) shows that the Co3-Co4 and Co7-Co8 "stripes" magnetic moments are coupled ferromagnetically and antiferromagnetically to their neighbors along the stripe direction in an "up-up-down-down" pattern. At higher temperatures up to ~60 K (near the magnetization hump), the magnetic moments rotate to form a zigzag pattern. Above ~60 K (well above the magnetization hump), the magnetic moments continue their rotation to align along the $[1\bar{1}0]$ direction. The magnitude and direction of the magnetic moment of the Co5 - Co6 pairs change non-monotonically as a function of temperature.

Fig. 13 presents the Co magnetic moments at 12 K refined as a function of Lu content. For x = 0.0, magnetic moment magnitudes of 3.19(6), 2.71(6), 2.69(8) and 2.60(9) $\mu_B$ were obtained for the Co1-Co2, Co3-Co4, Co5-Co6 and Co7-Co8 pairs, respectively. The smaller than expected magnetic moments (3 $\mu_B$ and 4 $\mu_B$ for high spin $Co^{2+}$ and $Co^{3+}$ states, respectively) may be explained by crystal field quenching effects and the persistence of significant and continuously enhanced short-range magnetic correlations with increasing Lu. Similarly, lower than expected magnetic moment magnitudes were observed by Khalyavin *et al.* [24] at 5 K and explained by the coexistence of diffuse scattering arising from fractional disorder of the magnetic moments together with the long-range ordered magnetic Co sublattice. The magnetic moments of Co1-Co2, Co3-Co4 and Co7-Co8 decrease linearly as a function of increased Lu content. Extrapolating the linear fits show that vanishing magnetic moments would be achieved for x between 0.8 and 0.9 consistent with the weak and broad neutron diffraction magnetic peaks disappearing at the same x values as shown in Fig. 3(b). The Co5-Co6 magnetic moment behavior, however, remains roughly constant for the substituted samples.

Figure 14 shows the Co magnetic moments as a function of temperature for x = 0.1 and 0.2. The overall features of the magnetic moment behaviors are similar for both samples, despite the larger scatter observed for x = 0.2 due to weaker long-range magnetic moment magnitudes.

## IV. Conclusions

This study establishes the impact of Lu substitution on the magnetic and structural properties of the geometrically frustrated $YBaCo_4O_7$ antiferromagnet. The full phase diagram of $Y_{1-x}Lu_xBaCo_4O_7$ is mapped out through temperature-dependent high-resolution neutron powder diffraction and magnetization measurements. Lu-substitution reduces the orthorhombic P*bn*$2_1$ phase transition temperature from 310 K (x = 0.0) to 110 K (x = 1.0). On the Lu side of the phase diagram (with x ≥ 0.7), a metastable monoclinic C*c* phase coexists with the orthorhombic P*bn*$2_1$ phase. The phase fraction of the monoclinic C*c* phase decreases with increasing Lu content, correlating with the weakening peak in the magnetization of these samples. The unit cell parameters *a* and *b*, and the volume decrease linearly with increasing Lu content due to the smaller ionic radius of Lu compared to Y while the *c*-axis is minimally affected.

Long-range AFM order observed below 110 K in YBaCo$_4$O$_7$ diminishes rapidly with increasing Lu content, leaving only short-range correlations for x > 0.5. Magnetic structures for all samples with x ≤ 0.5 were determined, and changes in the magnetic moments were correlated with the cusp/hump feature observed in the magnetization data of the Y-rich compositions. The refinements demonstrate the antiferromagnetic alignment of specific Co atom pairs in the *ab* plane in both the Kagomé and triangular layers. Some moments in the Kagomé layers align nearly parallel giving the appearance of 'stripes' along one of the *ab* plane diagonals. The behavior of the Co magnetic moments, associated with composition and temperature, appears to depend on the strong competition between LRO and SRO magnetic ordering.

**Acknowledgements**

This work was supported by the US Department of Energy, Office of Science, Basic Energy Sciences, Materials Science and Engineering Division. Sevda Sahinbay would like to thank Scientific and Technological Council of Turkey (TUBITAK) for financially supporting her through International Research Fellowship Program (2219) during the completion of this work. A portion of this research used resources at the Spallation Neutron Source, a DOE Office of Science User Facility operated by the Oak Ridge National Laboratory.

# References

[1] S. Nakatsuji, Y. Nambu, H. Tonomura, O. Sakai, S. Jonas, C. Broholm, H. Tsunetsugu, Y. Qiu, and Y. Maeno, *Spin Disorder on a Triangular Lattice*, Science (1979) **309**, 1697 (2005).

[2] M. F. Collins and O. A. Petrenko, Review/Synthèse*: Triangular Antiferromagnets*, Can J Phys **75**, 605 (1997).

[3] T. Katsufuji, S. Mori, M. Masaki, Y. Moritomo, N. Yamamoto, and H. Takagi, *Dielectric and Magnetic Anomalies and Spin Frustration in Hexagonal RMnO$_3$ (R = Y, Yb, and Lu)*, Phys Rev B **64**, 104419 (2001).

[4] J. N. Reimers, A. J. Berlinsky, and A.-C. Shi, *Mean-Field Approach to Magnetic Ordering in Highly Frustrated Pyrochlores*, Phys Rev B **43**, 865 (1991).

[5] X.-Y. Wang, L. Wang, Z.-M. Wang, and S. Gao, *Solvent-Tuned Azido-Bridged Co$^{2+}$ Layers: Square, Honeycomb, and* Kagomé, J Am Chem Soc **128**, 674 (2006).

[6] R. Moessner and S. L. Sondhi, *Ising Models of Quantum Frustration*, Phys Rev B **63**, 224401 (2001).

[7] E. Mengotti, L. J. Heyderman, A. F. Rodríguez, F. Nolting, R. V. Hügli, and H.-B. Braun, *Real-Space Observation of Emergent Magnetic Monopoles and Associated Dirac Strings in Artificial Kagome Spin Ice*, Nat Phys **7**, 68 (2011).

[8] R. Moessner and J. T. Chalker, *Properties of a Classical Spin Liquid: The Heisenberg Pyrochlore Antiferromagnet*, Phys Rev Lett **80**, 2929 (1998).

[9] S.-H. Lee, C. Broholm, W. Ratcliff, G. Gasparovic, Q. Huang, T. H. Kim, and S.-W. Cheong, *Emergent Excitations in a Geometrically Frustrated Magnet*, Nature **418**, 856 (2002).

[10] S. T. Bramwell and M. J. P. Gingras, *Spin Ice State in Frustrated Magnetic Pyrochlore Materials*, Science (1979) **294**, 1495 (2001).

[11] M. J. Harris, S. T. Bramwell, D. F. McMorrow, T. Zeiske, and K. W. Godfrey, *Geometrical Frustration in the Ferromagnetic Pyrochlore Ho$_2$Ti$_2$O$_7$*, Phys Rev Lett **79**, 2554 (1997).

[12] S. Avci, O. Chmaissem, H. Zheng, A. Huq, D. D. Khalyavin, P. W. Stephens, M. R. Suchomel, P. Manuel, and J. F. Mitchell, *Kinetic Control of Structural and Magnetic States in LuBaCo$_4$O$_7$*, Phys Rev B **85**, 094414 (2012).

[13] L. C. Chapon, P. G. Radaelli, H. Zheng, and J. F. Mitchell, *Competing Magnetic Interactions in the Extended Kagomé System YBaCo$_4$O$_7$*, Phys Rev B **74**, 172401 (2006).

[14] M. Soda, K. Morita, G. Ehlers, F. Ye, T. Tohyama, H. Yoshizawa, T. Masuda, and H. Kawano-Furukawa, *Magnetic Diffuse Scattering of YBaCo$_4$O$_7$ and LuBaCo$_4$O$_7$ on Kagome and Triangular Lattices*, J Physical Soc Japan **90**, 074704 (2021).

[15] T. Sarkar, M. M. Seikh, V. Pralong, V. Caignaert, and B. Raveau, *Magnetism of the "114" Orthorhombic Charge Ordered CaBaCo$_4$O$_7$ Doped with Zn or Ga: A Spectacular Valency Effect*, J Mater Chem **22**, 18043 (2012).

[16] A. Huq, J. F. Mitchell, H. Zheng, L. C. Chapon, P. G. Radaelli, K. S. Knight, and P. W. Stephens, *Structural and Magnetic Properties of the Kagomé Antiferromagnet YbBaCo$_4$O$_7$*, J Solid State Chem **179**, 1136 (2006).

[17] N. Nakayama, T. Mizota, Y. Ueda, A. N. Sokolov, and A. N. Vasiliev, *Structural and Magnetic Phase Transitions in Mixed-Valence Cobalt Oxides REBaCo$_4$O$_7$ (RE=Lu, Yb, Tm)*, J Magn Magn Mater **300**, 98 (2006).

[18] A. Maignan, V. Caignaert, V. Pralong, D. Pelloquin, and S. Hébert, *Impact of Metal Substitutions for Cobalt in YBaCo$_4$O$_7$*, J Solid State Chem **181**, 1220 (2008).

[19] T. Sarkar, V. Caignaert, V. Pralong, and B. Raveau, *Hysteretic "Magnetic-Transport-Structural" Transition in "114" Cobaltites: Size Mismatch Effect*, Chemistry of Materials **22**, 6467 (2010).

[20] M. Y. Ruan, Z. W. Ouyang, Y. M. Guo, Y. C. Sun, J. J. Cheng, Z. C. Xia, and G. H. Rao, *Spin-Glass-like Freezing in Geometrically Frustrated Compound InBaCo$_4$O$_7$*, Solid State Sci **45**, 1 (2015).

[21] V. Caignaert, V. Pralong, A. Maignan, and B. Raveau, *Orthorhombic Kagome Cobaltite CaBaCo$_4$O$_7$: A New Ferrimagnet with a $T_C$ of 70 K*, Solid State Commun **149**, 453 (2009).

[22] Z. Qu, L. Ling, L. Zhang, L. Pi, and Y. Zhang, *Magnetic Properties of the Ferrimagnetic Cobaltite CaBaCo$_4$O$_7$*, Solid State Commun **151**, 917 (2011).

[23] P. Manuel, L. C. Chapon, P. G. Radaelli, H. Zheng, and J. F. Mitchell, *Magnetic Correlations in the Extended Kagome YBaCo$_4$O$_7$ Probed by Single-Crystal Neutron Scattering*, Phys Rev Lett **103**, 037202 (2009).

[24] D. D. Khalyavin, P. Manuel, B. Ouladdiaf, A. Huq, H. Zheng, J. F. Mitchell, and L. C. Chapon, *Spin-Ordering and Magnetoelastic Coupling in the Extended Kagome System YBaCo$_4$O$_7$*, Phys Rev B **83**, 094412 (2011).

[25] M. J. R. Hoch, P. L. Kuhns, S. Yuan, T. Besara, J. B. Whalen, T. Siegrist, A. P. Reyes, J. S. Brooks, H. Zheng, and J. F. Mitchell, *Evidence for an Internal-Field-Induced Spin-Flop Configuration in the Extended Kagome YBaCo$_4$O$_7$*, Phys Rev B **87**, 064419 (2013).

[26] A. Huq, M. Kirkham, P. F. Peterson, J. P. Hodges, P. S. Whitfield, K. Page, T. Hugle, E. B. Iverson, A. Parizzia, and G. Rennichb, *POWGEN: Rebuild of a Third-Generation Powder Diffractometer at the Spallation Neutron Source*, J Appl Crystallogr **52**, 1189 (2019).

[27] H. M. Rietveld, *A Profile Refinement Method for Nuclear and Magnetic Structures*, J Appl Crystallogr **2**, 65 (1969).

[28] B. H. Toby and R. B. Von Dreele, *GSAS-II: The Genesis of a Modern Open-Source All Purpose Crystallography Software Package*, J Appl Crystallogr **46**, 544 (2013).

[29] S. Lee, W. Lee, K. J. Lee, B. J. Kim, B. J. Suh, H. Zheng, J. F. Mitchell, and K. Y. Choi, *Muon Spin Relaxation Study of Spin Dynamics in the Extended Kagome Systems YBaCo$_4$O$_{7+\delta}$ ($\delta = 0, 0.1$)*, Phys Rev B **97**, 104409 (2018).

[30] *See Supplemental Material at [URL Will Be Inserted by Publisher] for Additional Information on Structures, Magnetic Properties and Neutron Powder Diffraction*, (unpublished).

[31] L. Vegard, *Die Konstitution Der Mischkristalle Und Die Raumfüllung Der Atome*, Zeitschrift Für Physik 1921 5:1 **5**, 17 (1921).

[32] R. D. Shannon, *Revised Effective Ionic Radii and Systematic Studies of Interatomic Distances in Halides and Chalcogenides*, Acta Crystallographica Section A **32**, 751 (1976).

**Table 1.** Atom positions of cobalt (Co) atoms in the magnetic phase at 12 K for the x = 0.1 sample.

| Atom | Color | x | y | z |
| --- | --- | --- | --- | --- |
| Co1 | Dark blue | 0.0015(24) | 0.0030(8) | -0.0609(17) |
| Co2 | Dark Blue | 0.4985(24) | 0.5030(8) | -0.0609(17) |
| Co3 | Green | 0.0017(19) | 0.1743(9) | -0.3159(11) |
| Co4 | Green | 0.4983(19) | 0.6743(9) | -0.3159(11) |
| Co5 | Purple | 0.2602(19) | 0.0904(9) | -0.8105(11) |
| Co6 | Purple | 0.2464(19) | 0.4179(9) | -0.3172(11) |
| Co7 | Cyan | 0.2398(20) | 0.5904(10) | -0.8105(11) |
| Co8 | Cyan | 0.2536(20) | 0.9179(10) | -0.3172(11) |

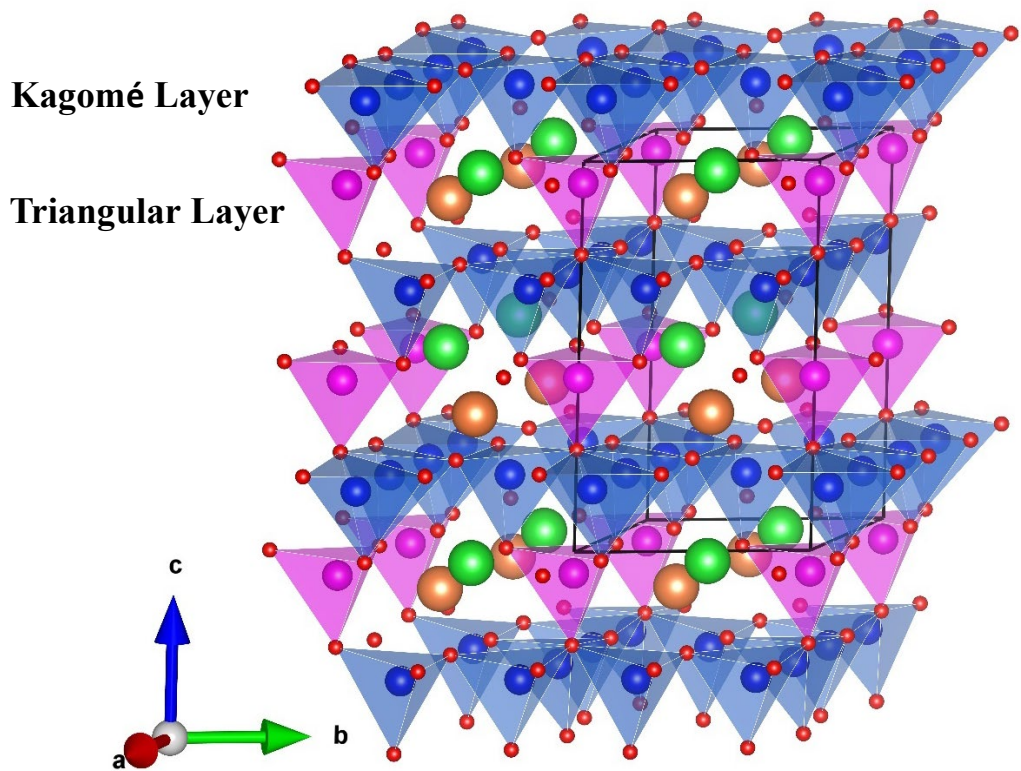

**Fig. 1.** Crystal structure of the $R$BaCo$_4$O$_7$ family with the trigonal symmetry of space group P31$c$. Cobalts atoms occupy the centers of the polyhedra shown in the Kagomé and triangular layers. Oxygen, barium, and $R$ atoms are represented by the red, green and orange spheres, respectively.

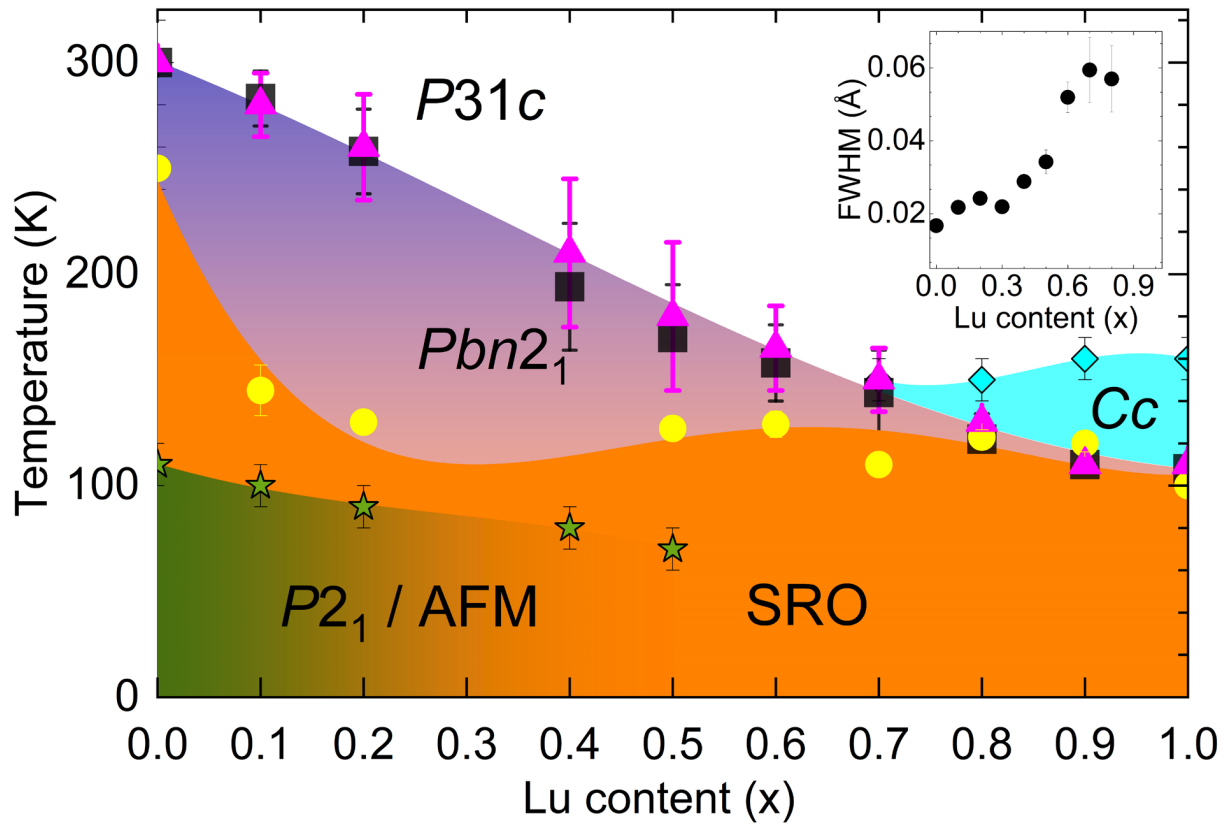

**Fig. 2.** Structural and magnetic phase diagram of $Y_{1-x}Lu_xBaCo_4O_7$ ($\Delta x = 0.1$). The black squares indicate the structural transition temperatures obtained from the first derivatives of the magnetization vs temperature curves, with error bars corresponding to the width of the peak on the first derivatives. The magenta triangles represent the structural transition temperatures ($T_{s1}$) to orthorhombic P$bn2_1$, obtained from neutron diffraction data. The error bars on the magenta triangles indicate the biphasic region where trigonal and orthorhombic phases coexist. The green stars denote the transition temperatures of long range ordered antiferromagnetic structures (LRO AFM) obtained from neutron diffraction data. Cyan diamonds mark the structural transition temperatures ($T_{s2}$) to the monoclinic C$c$ phase. The error bars on cyan diamonds and green stars are set to 10 K, as the neutron data were collected with 10K intervals. Yellow circles represent the onset of short-range magnetic correlations (SRO), with data for x = 0.0 and x = 1.0 taken from Refs [12,13], respectively. The error bars on yellow circles for $0.1 \leq x \leq 0.9$ are determined from the Boltzmann function fit to the integrated intensity of the (012) magnetic reflection (see Fig. 3). The borders of the violet (orthorhombic P$bn2_1$ phase), cyan (C$c$ phase), green (AFM) and orange (SRO) regions are guides to the eye obtained by polynomial fits to the corresponding data points from neutron diffraction. Neutron diffraction data were collected on warming after cooling with a rate of ∼3.5 K/min. The phase diagram represents the equilibrium condition where the P$bn2_1$ phase is the stable phase below $T_{s1}$ for $x \geq 0.7$. Inset: Full width at half maximum (FWHM) of the (012) magnetic peak as a function of Lu content. Neutron diffraction data for the x=0.0 sample are collected at 5 K, for the rest of the samples at 12 K.

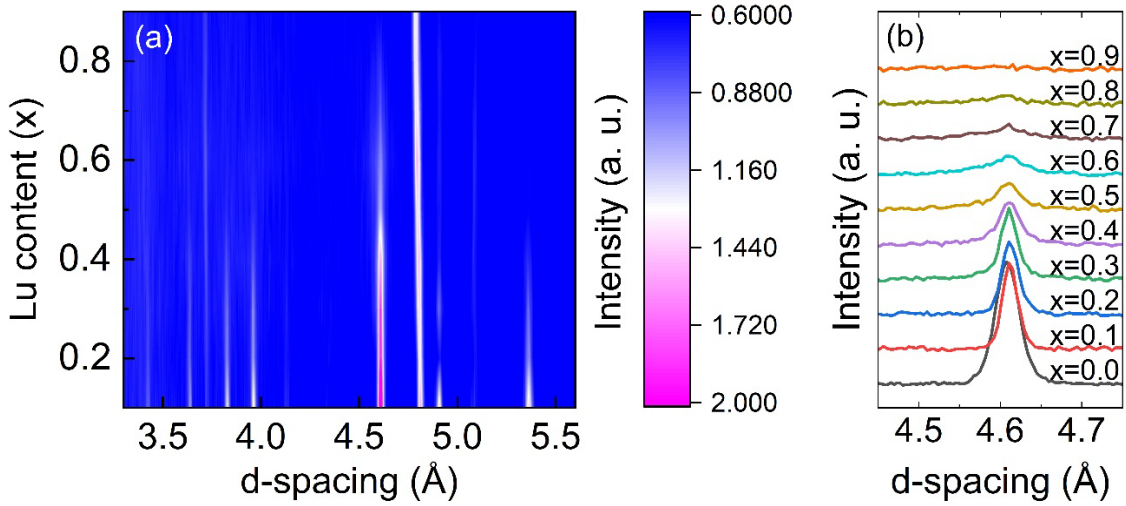

**Fig. 3.** (a) Contour plot displaying a portion of the neutron diffraction data for $Y_{1-x}Lu_xBaCo_4O_7$ and highlighting the suppression of magnetic peaks with increasing Lu content. (b) The 012 magnetic reflection for each sample at 5 K for x=0.0, 10 K for x=1.0 and 12 K for the other Lu substituted samples. The measurements were conducted after fast cooling at a rate of ∼3.5K/min, as outlined in the experimental details.

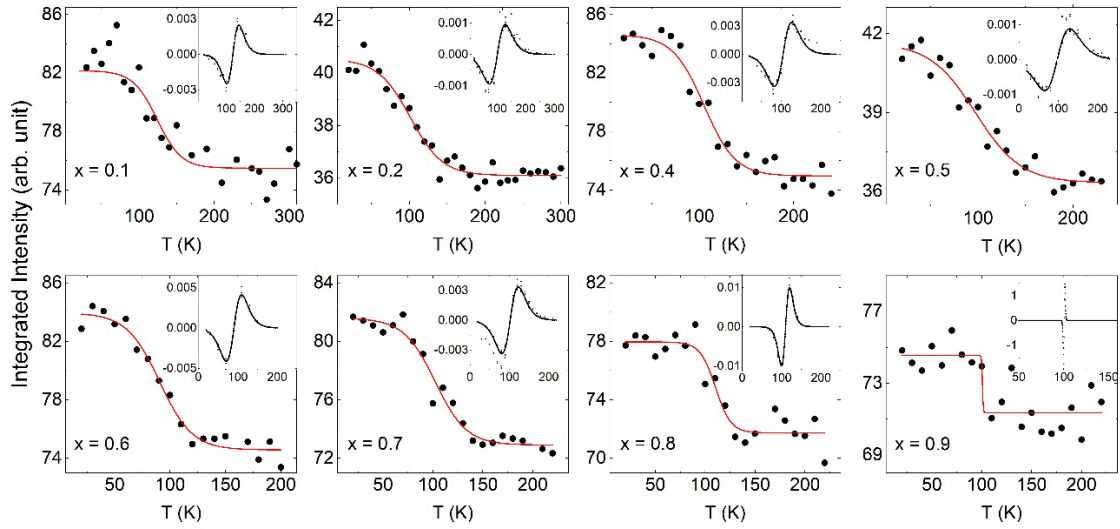

**Fig. 4.** Integrated intensities represented by black closed circles of the (012) magnetic reflection for samples with 0.1 ≤ x ≤ 0.9 samples. The solid red lines depict Boltzmann function fits to the data. Insets show the second derivatives of the Boltzmann function fits as a function of temperature. The neutron diffraction data were collected during the warming phase after fast cooling at a rate of approximately 3.5 K/min down to 12 K.

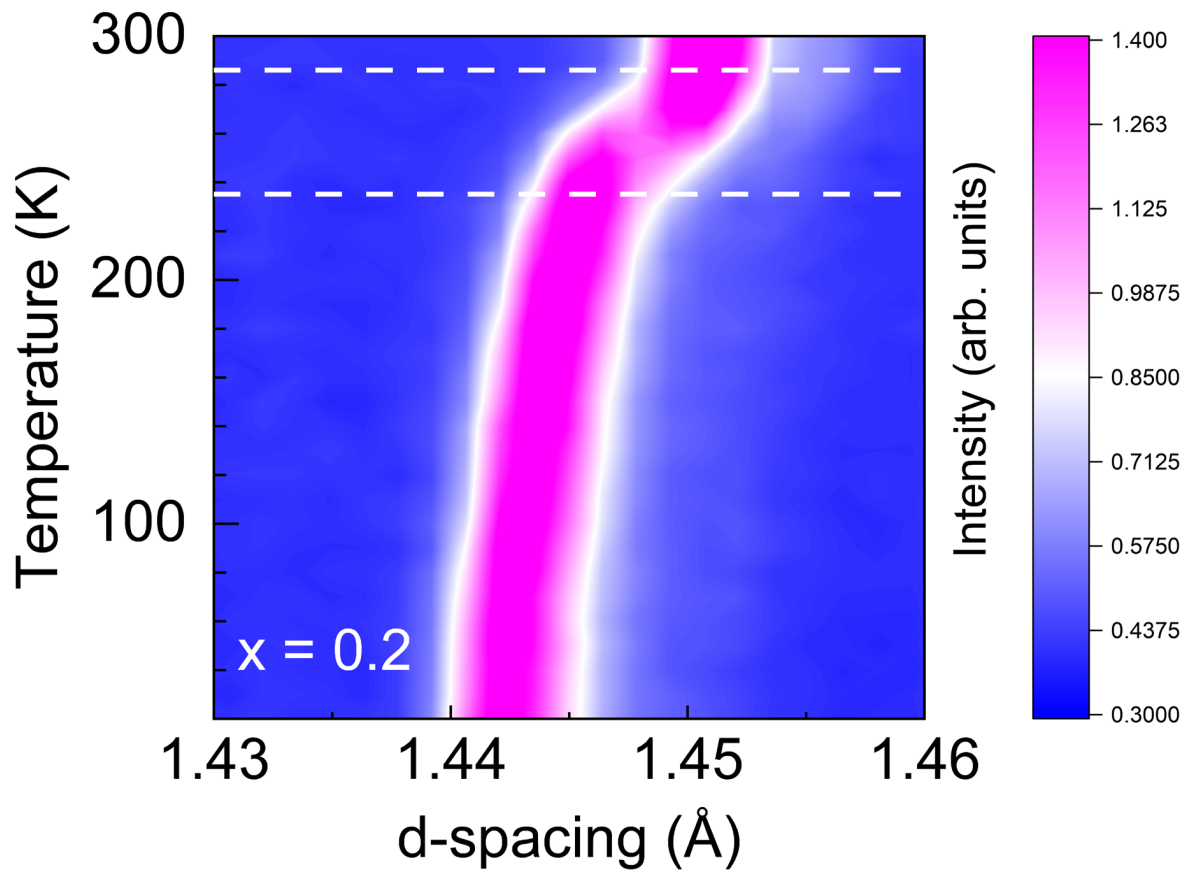

**Fig. 5.** A portion of neutron diffraction data for the x = 0.2 sample illustrating the structural transition from trigonal to orthorhombic phases. The biphasic region between the two horizontal dashed lines indicates the coexistence of both phases. The neutron diffraction data were collected during the warming phase after fast cooling at a rate of approximately 3.5 K/min down to 12 K.

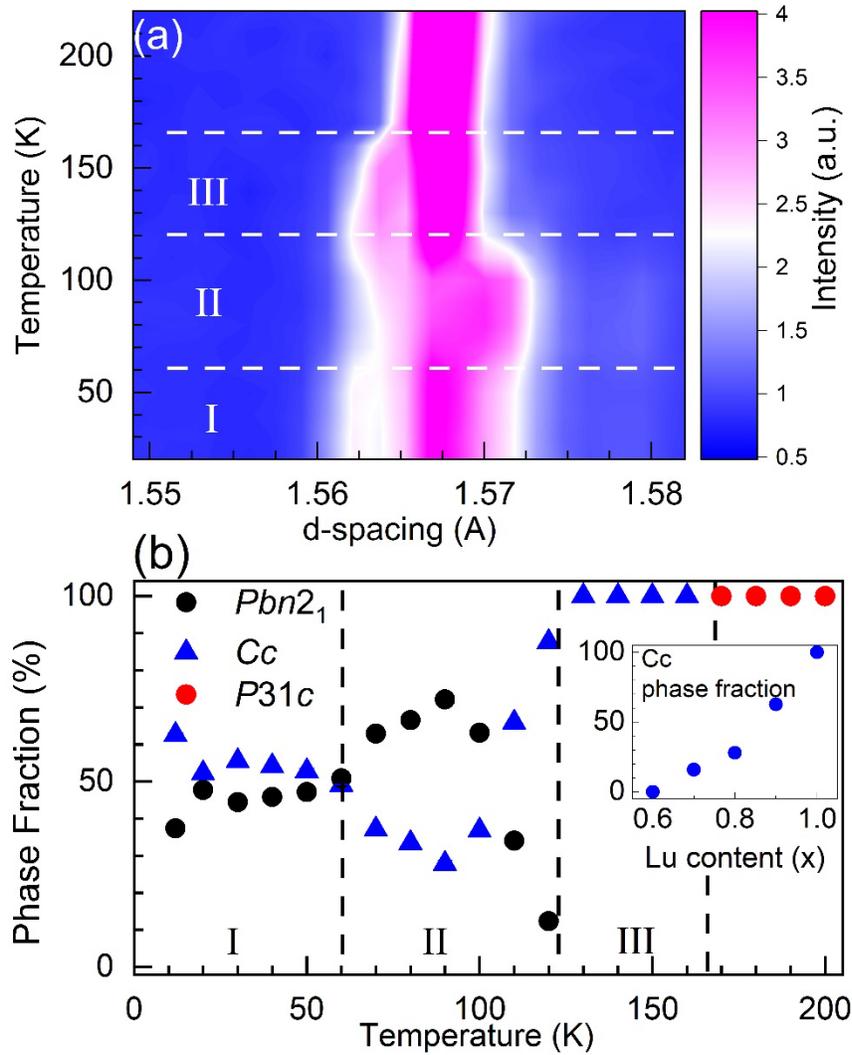

**Fig. 6.** (a) A contour plot displaying a portion of neutron diffraction data for the x = 0.9 sample. (b) Calculated phase fractions of P$bn2_1$ and C$c$ phases for the x=0.9 sample. Region III corresponds to a pure C$c$ phase, while Regions I and II indicate a mixture of P$bn2_1$ and C$c$ phases. In Region I, the C$c$ phase dominates, while in Region II, P$bn2_1$ is the dominant phase until it starts disappearing above 100 K. The inset displays the phase fraction of the C$c$ phase as a function of Lu content at 12 K. The weight fraction of the orthorhombic P$bn2_1$ phase exhibits an opposite trend. The neutron diffraction data were collected during the warming phase after fast cooling at a rate of approximately 3.5 K/min down to 12 K.

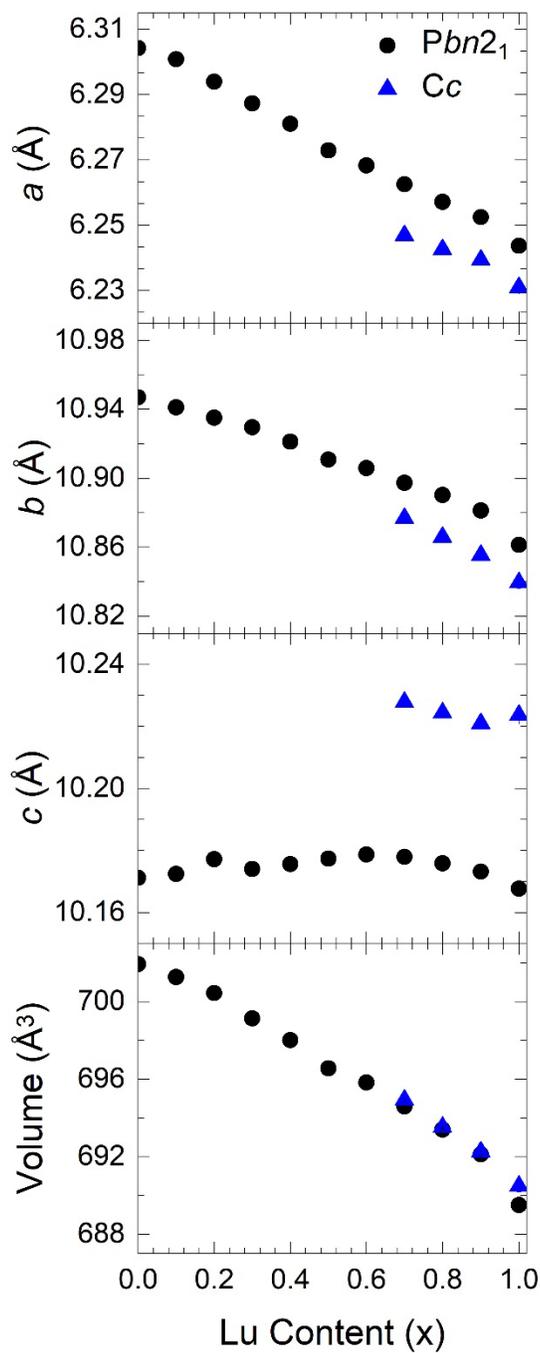

**Fig. 7.** The variation in unit cell parameters as a function of Lu content at 12 K. Data points for x = 1.0 are taken from Ref. [12]. Error bars, smaller than the symbols, indicate the uncertainties associated with the measurements. The neutron diffraction data were collected during the warming phase after fast cooling at a rate of approximately 3.5 K/min down to 12 K.

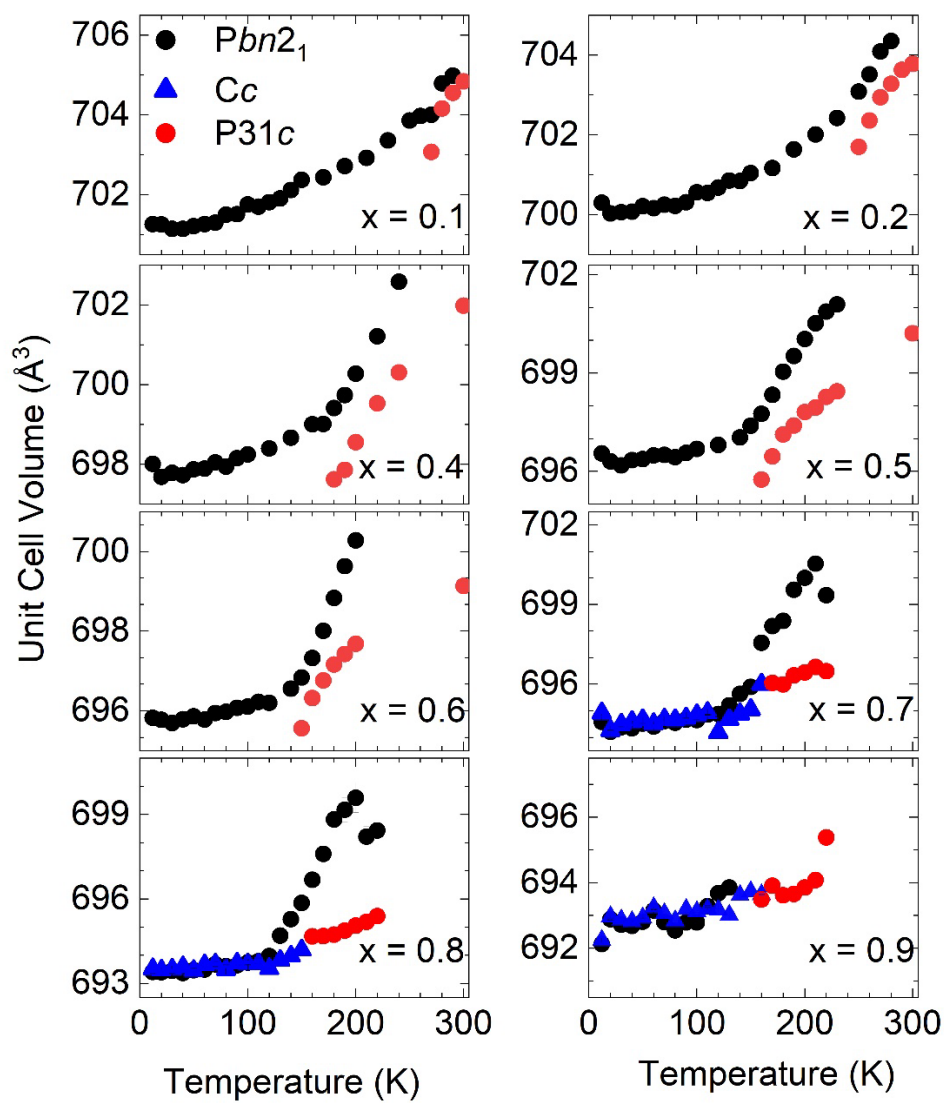

**Fig. 8.** Unit cell volume as a function of temperature for $Y_{1-x}Lu_xBaCo_4O_7$. The error bars, smaller than the symbols, represent the uncertainties associated with the measurements. The neutron diffraction data were collected on warming after fast cooling at a rate of approximately 3.5 K/min down to 12 K.

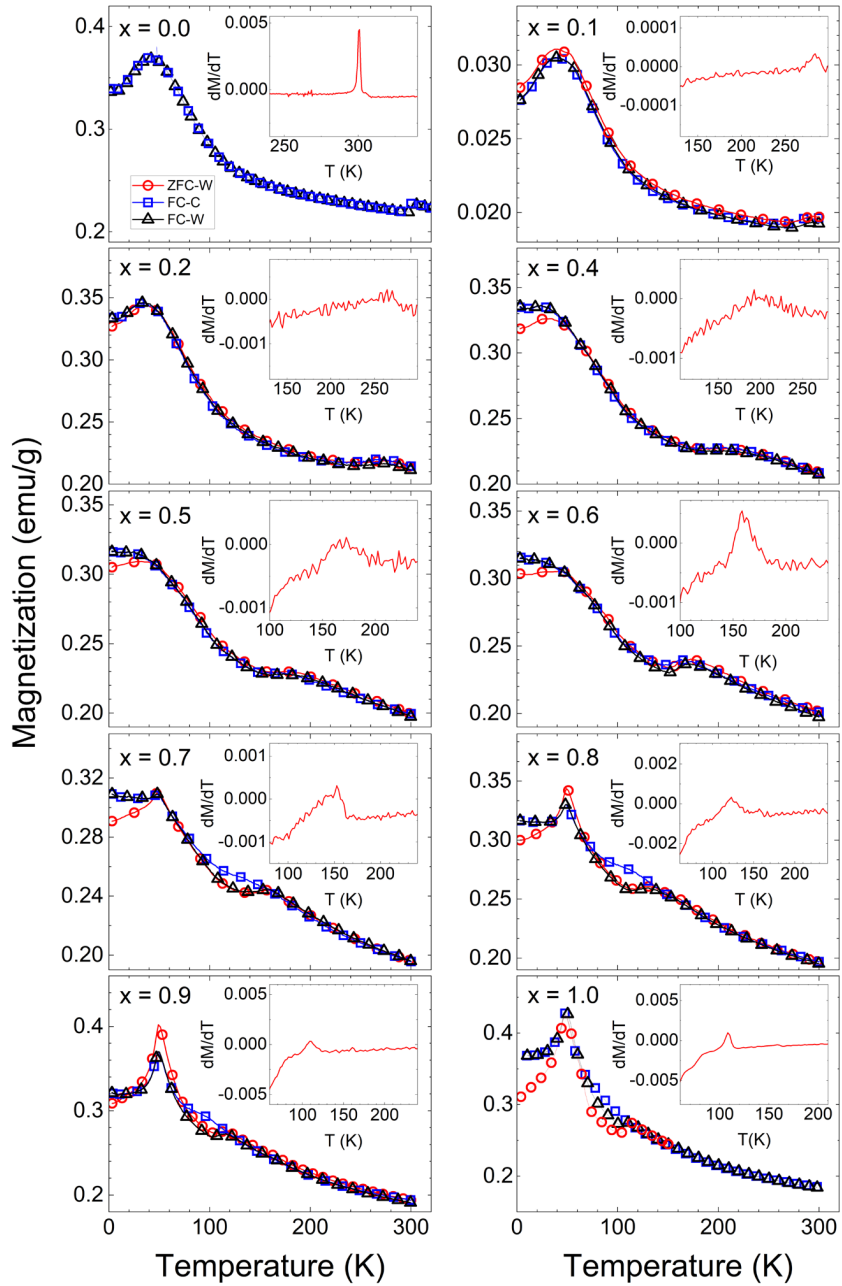

**Fig. 9.** Temperature dependence of magnetization for $Y_{1-x}Lu_xBaCo_4O_7$. The data were collected under a magnetic field of 1 T using various conditions: on warming after cooling in zero magnetic field (ZFC-W), on warming after cooling in a magnetic field (FC-W), and on cooling in a magnetic field (FC-C). The cooling and warming rate was set at 2 K/min. To enhance clarity, some data points are skipped. The insets show first-derivatives of the corresponding magnetization data.

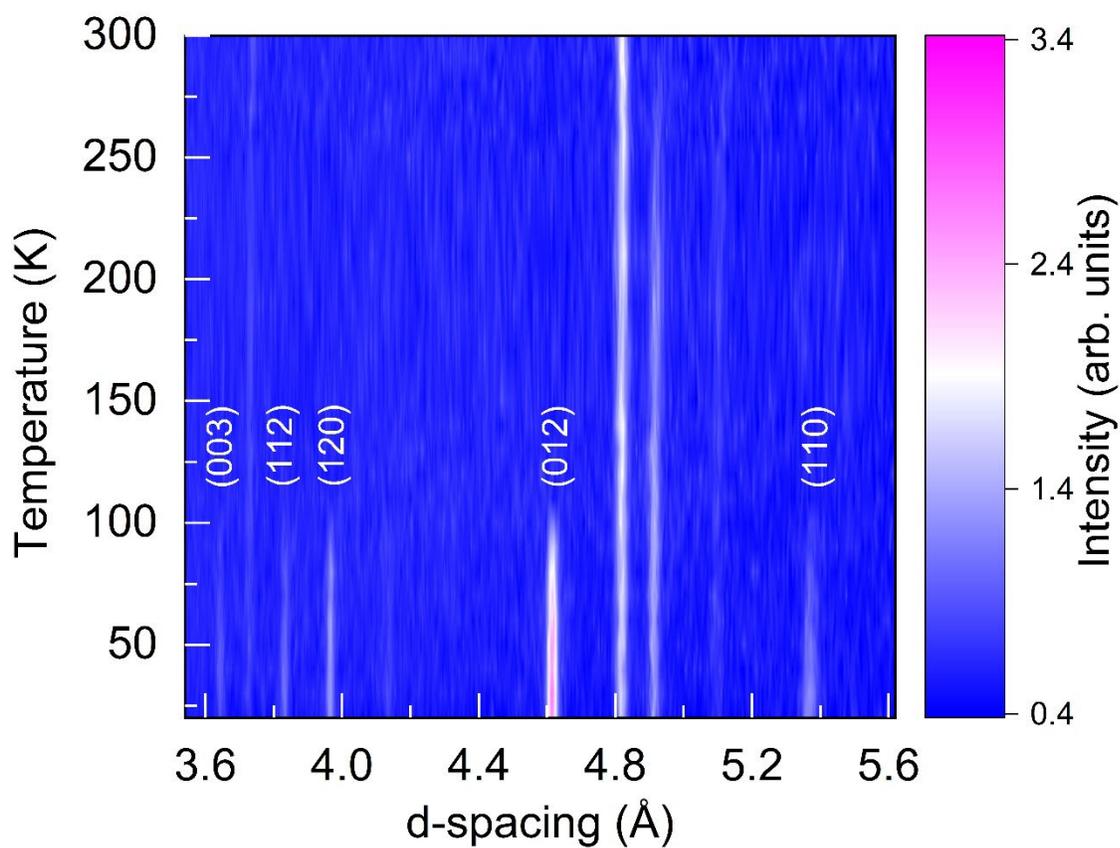

**Fig. 10.** Contour plot of neutron diffraction data for x = 0.1 sample as a function of temperature. Magnetic peaks (indexed) are observed below 100 K. The neutron diffraction data were collected on warming after fast cooling at a rate of approximately 3.5 K/min down to 12 K.

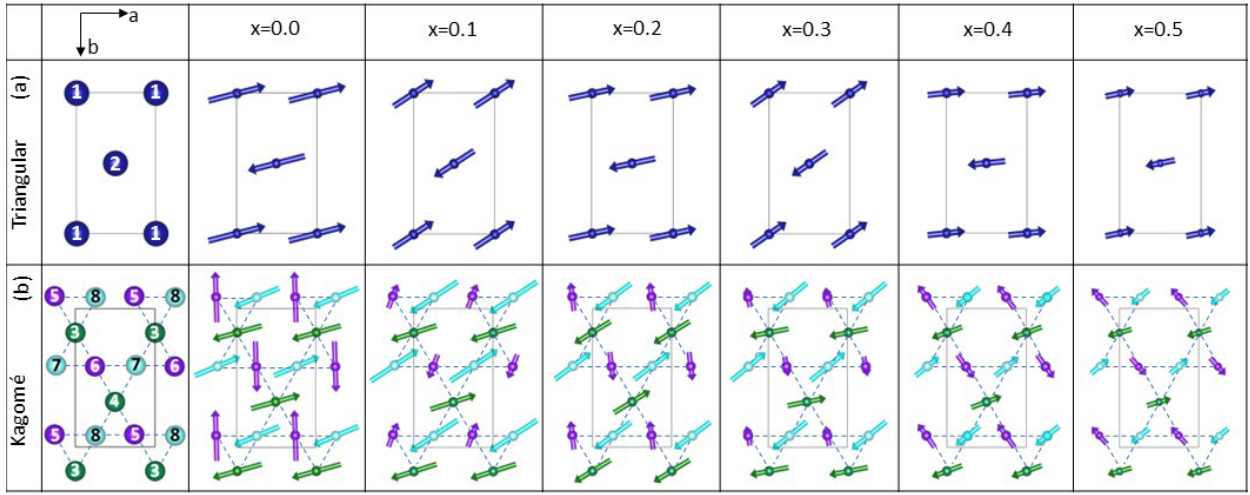

**Fig. 11.** Magnetic structures of long-range ordered $Y_{1-x}Lu_xBaCo_4O_7$ at 5 K for x = 0.0 and at 12 K for the Lu substituted samples in both the triangular layer (a) and Kagomé layer (b). In the triangular layer, the antiferromagnetically coupled Co1-Co2 atom pairs are colored blue, while in the Kagomé layer, Co3-Co4 pairs are green, Co5-Co6 pairs are purple, and Co7-Co8 pairs are cyan. The arrows represent the relative magnetic moments, all lying in the ab-plane.

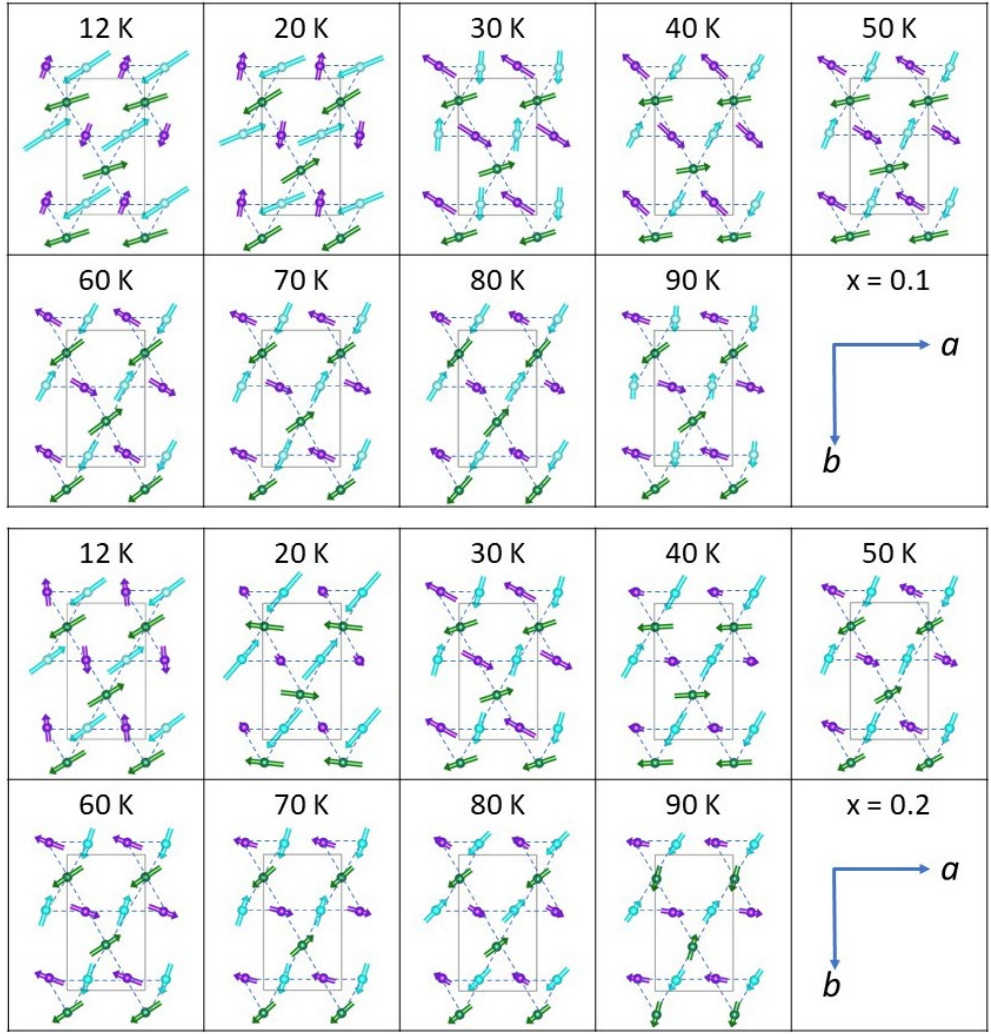

**Fig. 12.** Effects of the temperature on the magnetic structure of the Kagomé layer for the x = 0.1 and 0.2 samples. In these representations, atom pairs that align antiferromagnetically share the same color: Co3-Co4 pairs are green, Co5-Co6 pairs are purple, and Co7-Co8 pairs are cyan. The arrows indicate the relative magnetic moments, all of which lie in the ab plane based on the magnetic $P12_1'1$ symmetry. The neutron diffraction data were collected on warming after fast cooling (~3.5 K/min) down to 12 K.

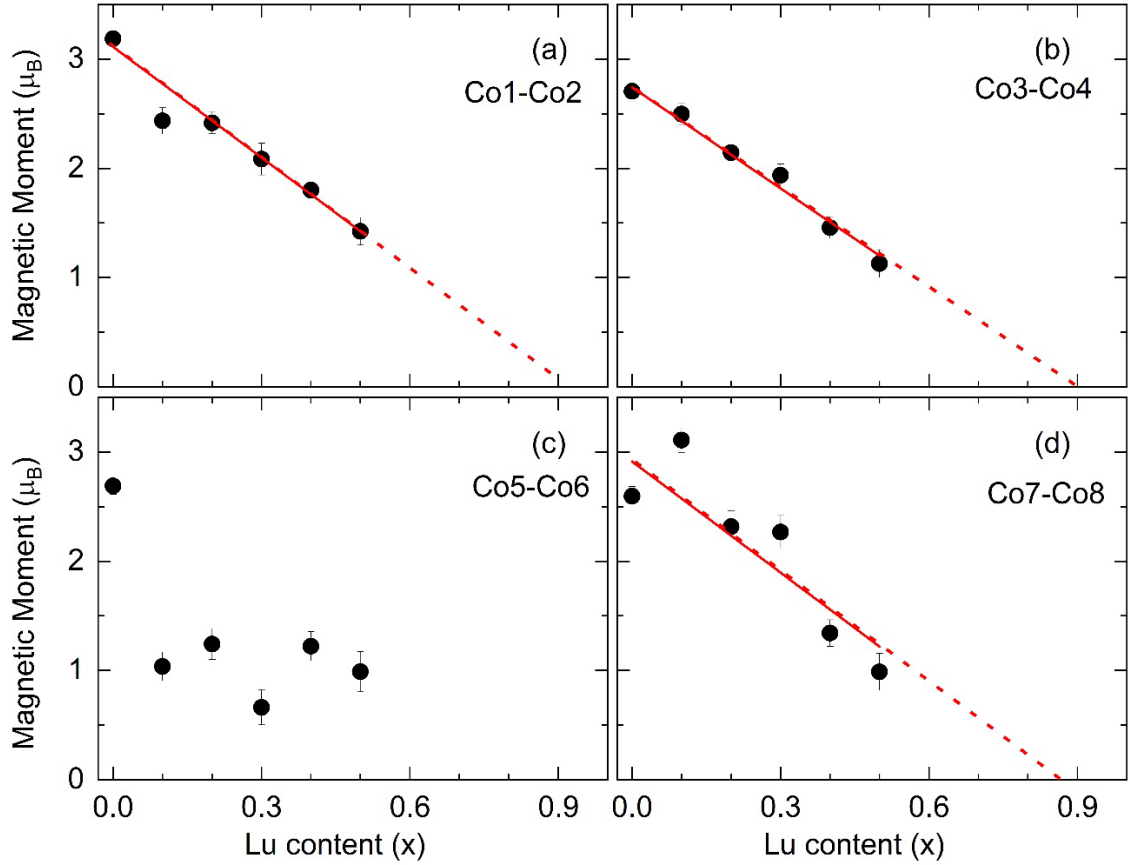

**Fig. 13.** Magnetic moments of Co1-Co2, Co3-Co4, Co5-Co6, Co7-Co8 as a function of Lu content. The red solid lines represent linear fits to the data, providing a visual guide. The red dotted lines show extrapolations of the linear fits. Neutron diffraction data were collected at 5 K for the x = 0.0 sample and at 12 K for all the Lu-substituted samples after fast cooling (~3.5 K/min).

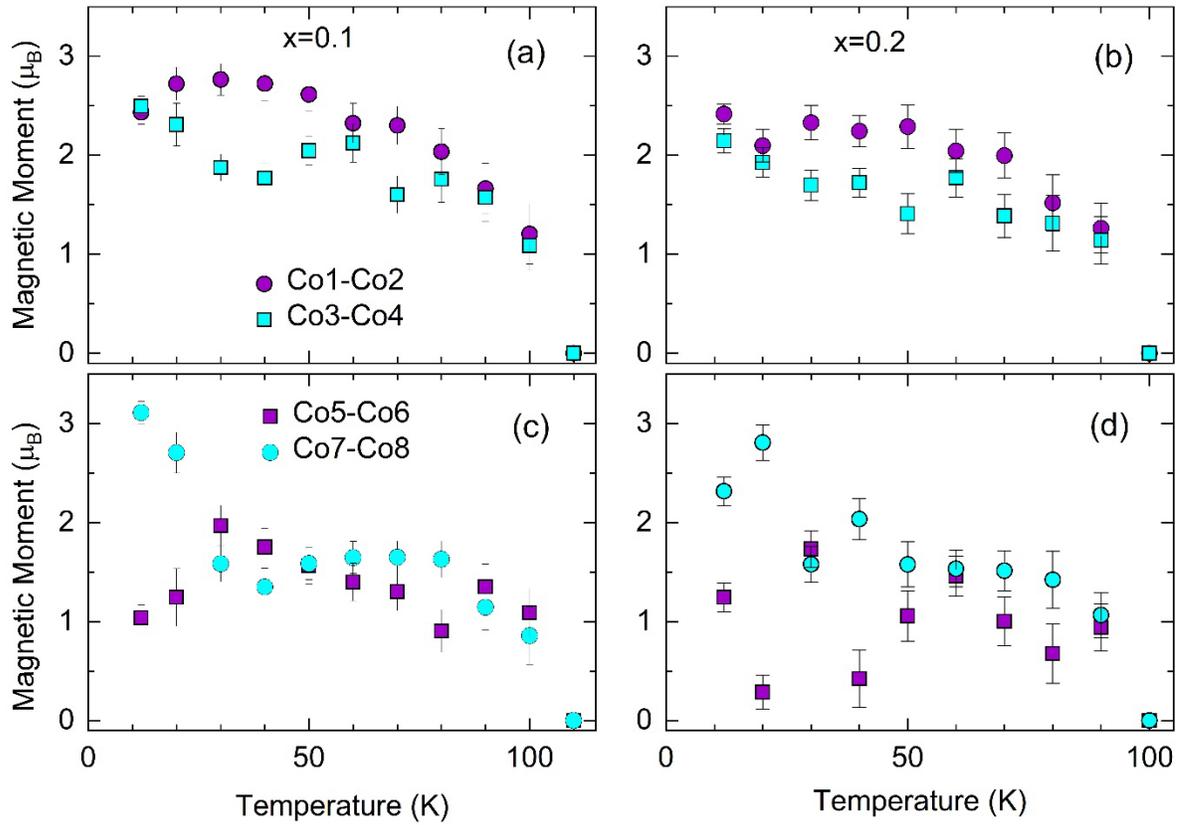

**Figure 14.** Magnetic moments of Co1-Co2 and Co3-Co4 (a) and (b) as well as Co5-Co6 and Co7-Co8 (c) and (d) as a function of temperature for x = 0.1 (left panels) and x = 0.2 (right panels). The neutron diffraction data were collected on warming after fast cooling (~3.5 K/min) down to 12 K.

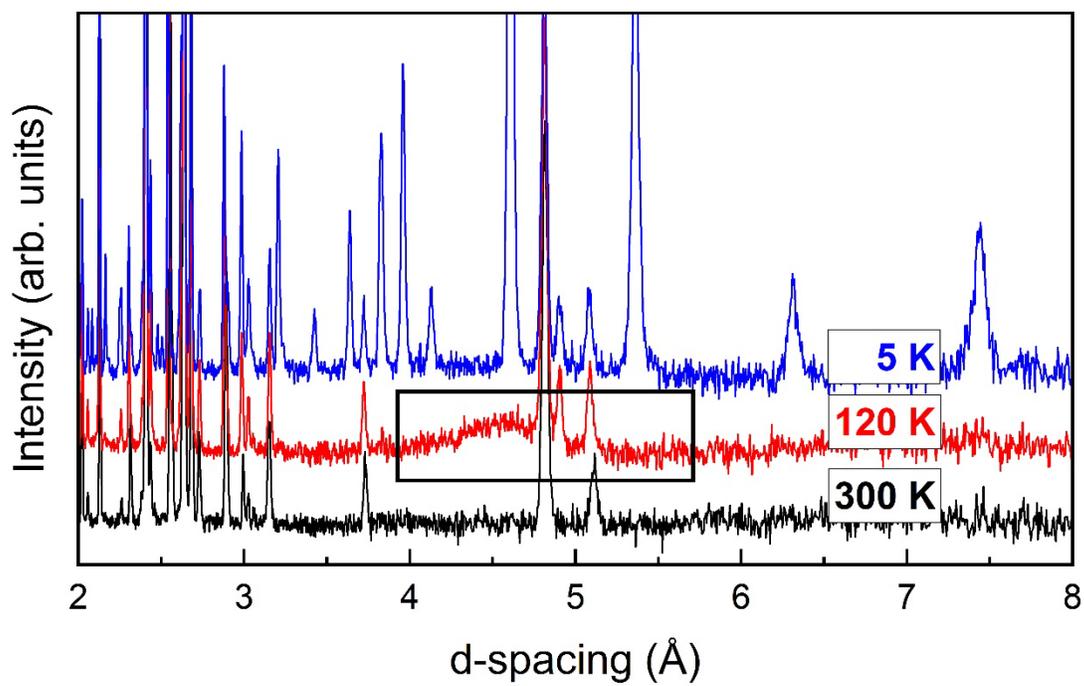

**Supplementary Figure 1** Neutron diffraction data for YBaCo$_4$O$_7$ at 5K, 120K, and 300K. Diffuse scattering region is highlighted with a rectangular box.

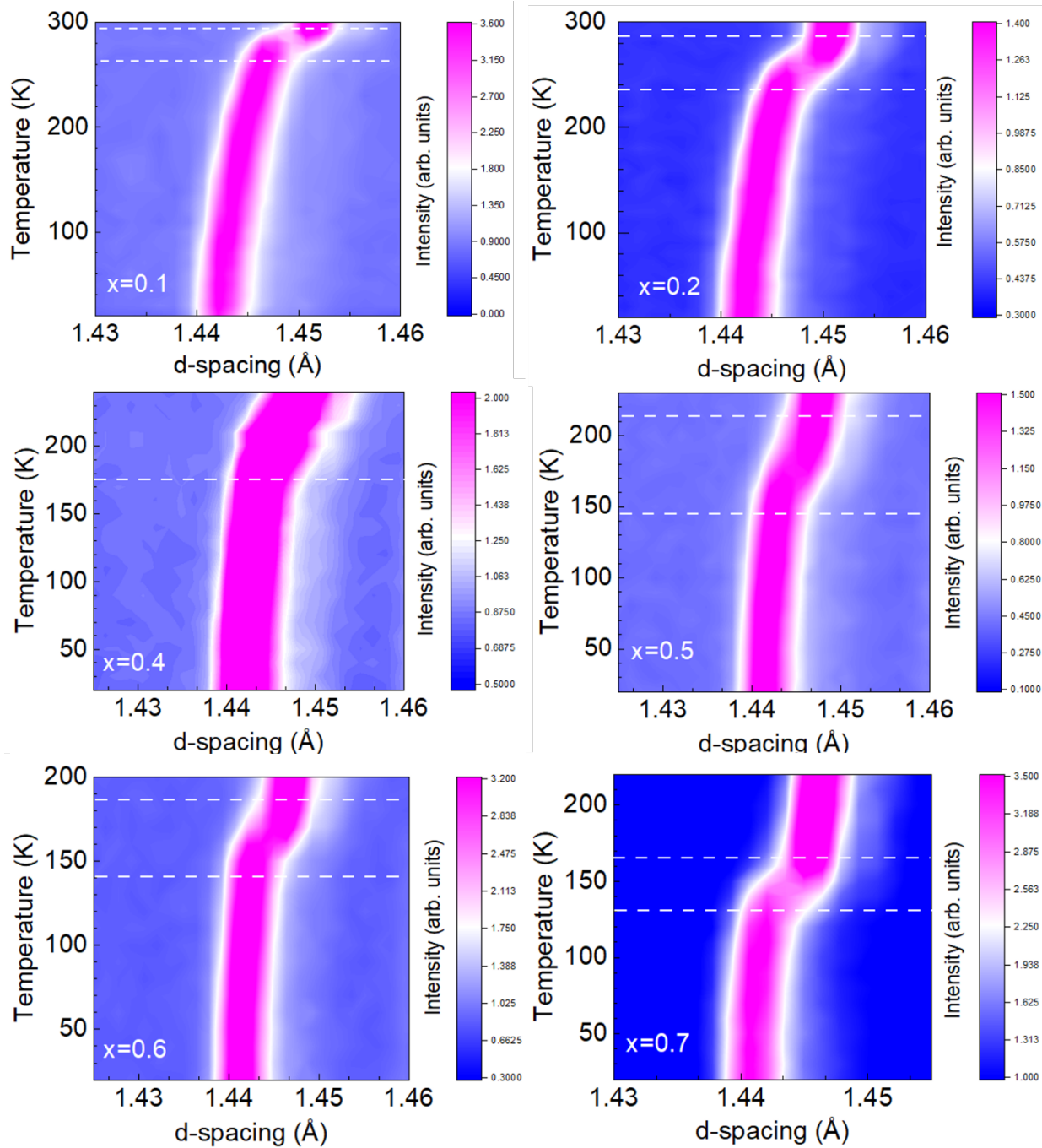

**Supplementary Figure 2.** Contour plot illustrating a portion of neutron diffraction data demonstrating the structural transition from trigonal to orthorhombic phases and the biphasic region where these two phases coexist. Neutron diffraction data were collected on warming following cooling with a rate of ~3.5 K/min.

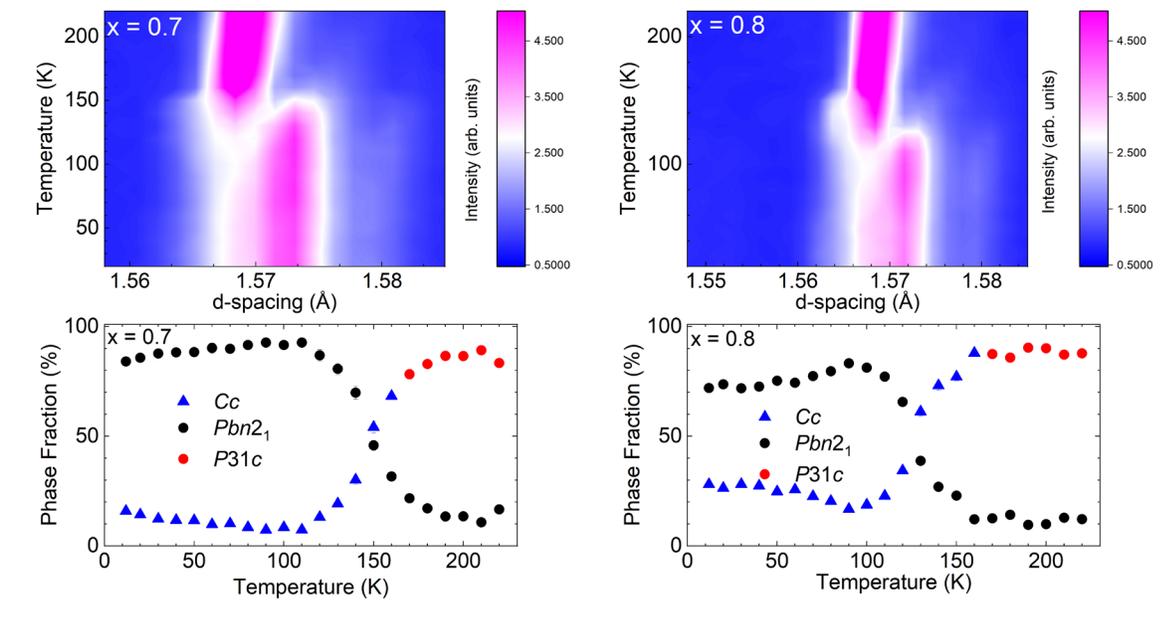

**Supplementary Figure 3.** Contour maps of partial neutron diffraction data and phase fractions of monoclinic, orthorhombic, and trigonal phases as a function of temperature for x = 0.7 (left panels) and x = 0.8 (right panels) samples. Neutron diffraction data were collected on warming after cooling with a rate of ~3.5 K/min.

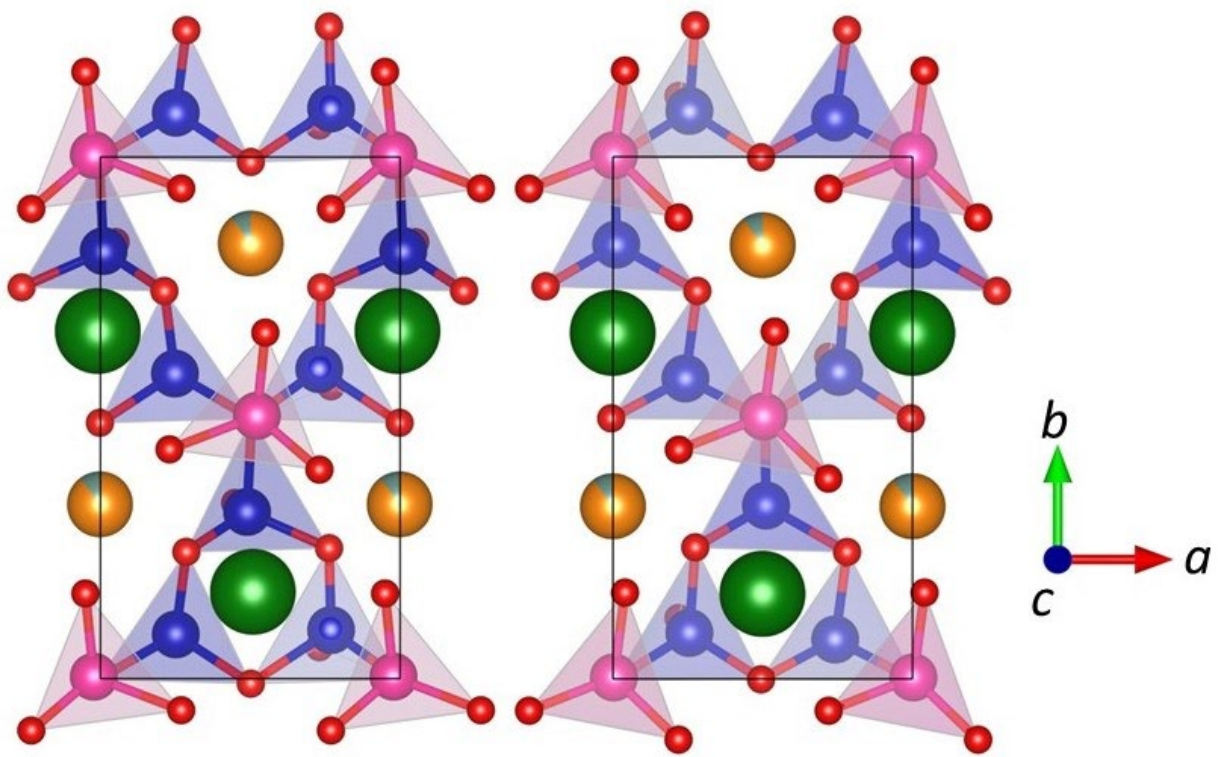

**Supplementary Figure 4.** Nuclear structures at 12 K of the P$bn2_1$ (left) and C$c$ (right) phases for the x = 0.9 sample. The blue and magenta, green, orange, and red spheres represent the Co, Ba, Lu/Y, and oxygen atoms, respectively.

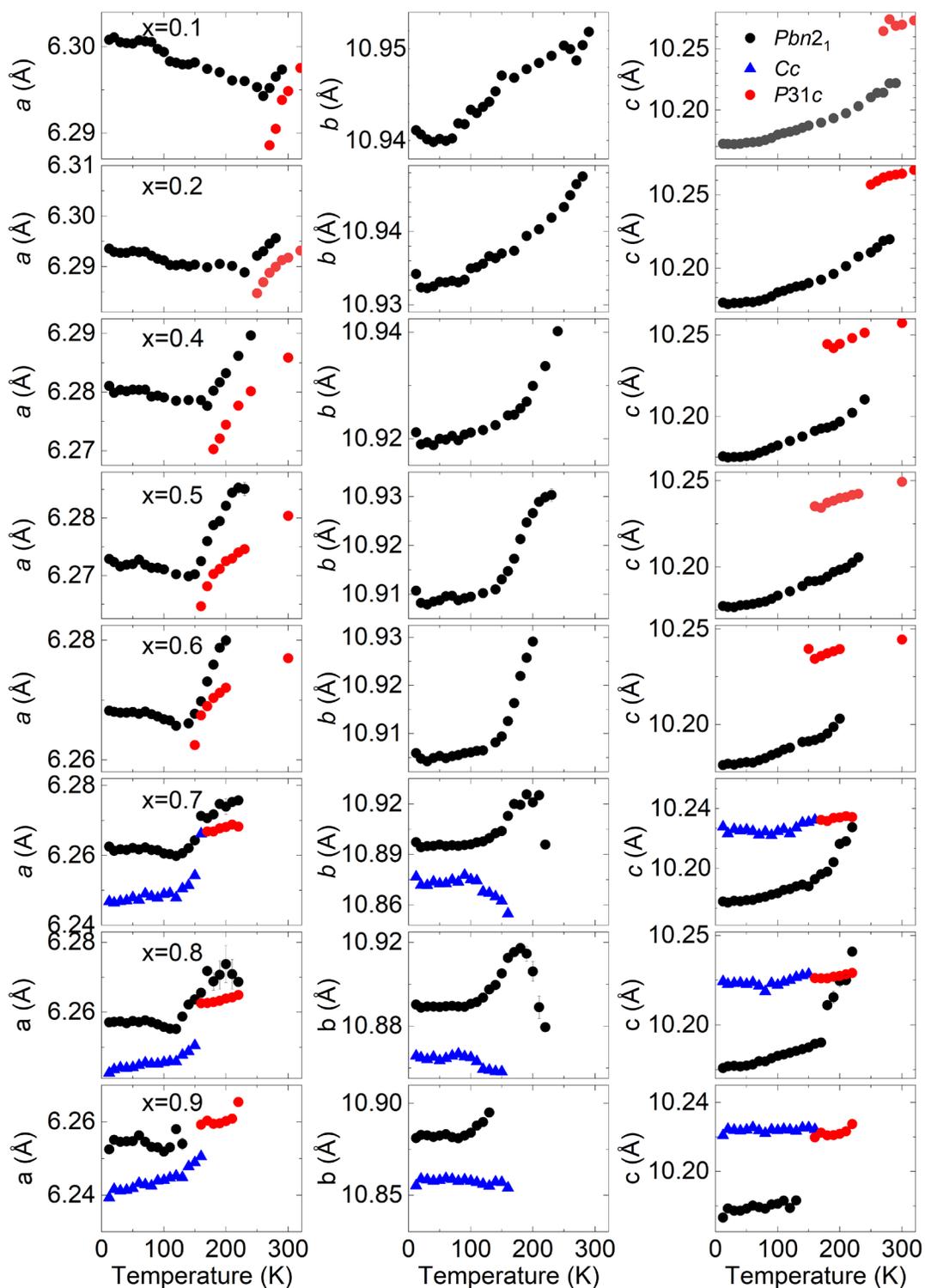

**Supplementary Figure 5.** Unit cell parameters as a function of temperature. The neutron diffraction data were collected on warming after fast cooling (∼3.5 K/min) down to 12 K.

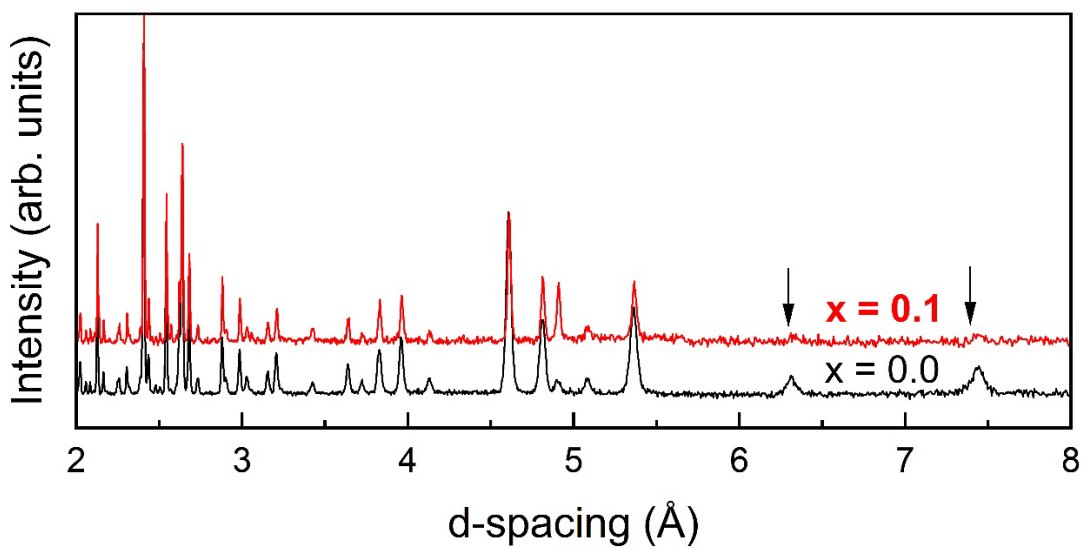

**Supplementary Figure 6.** Neutron diffraction data are presented for the x = 0.0 (5K) and x = 0.1 (12K) samples, emphasizing the presence of two additional magnetic peaks specifically in the x = 0.0 dataset. The neutron diffraction data were collected on warming the sample after rapid cooling at a rate of approximately 3.5 K/min down to 12 K.

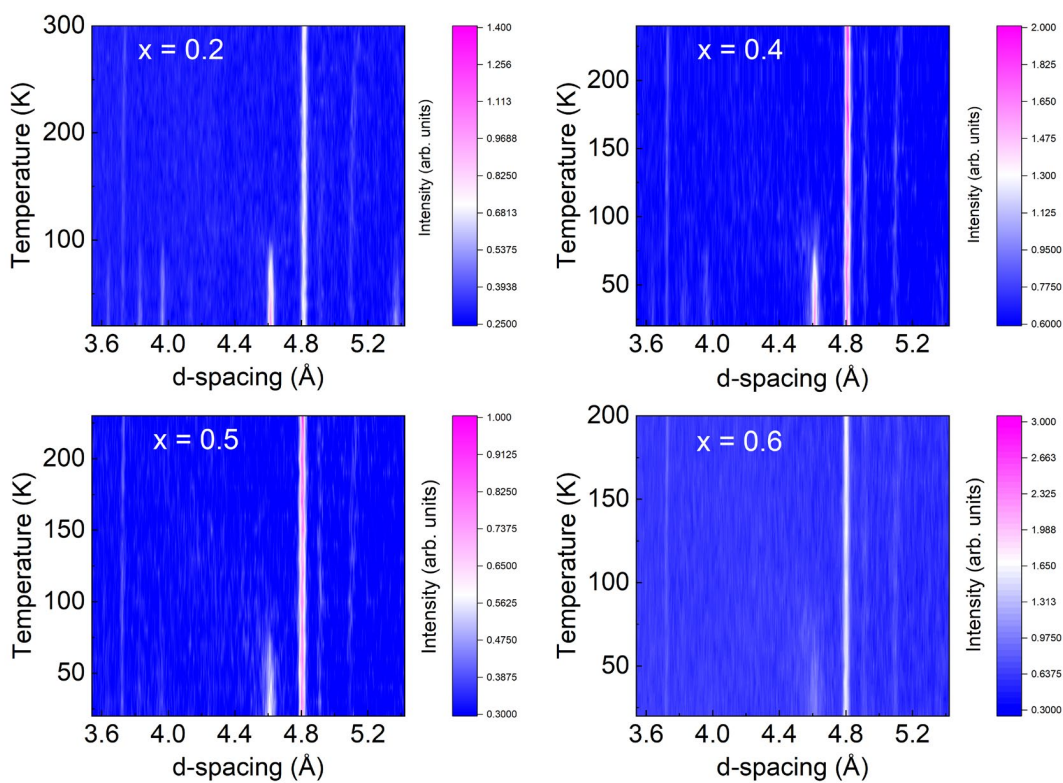

**Supplementary Figure 7.** Contour maps display the presence of magnetic peaks in neutron diffraction data portions for samples with x = 0.2, 0.4, 0.5, and 0.6. Data collection performed on warming following rapid cooling at approximately 3.5 K/min down to 12 K.

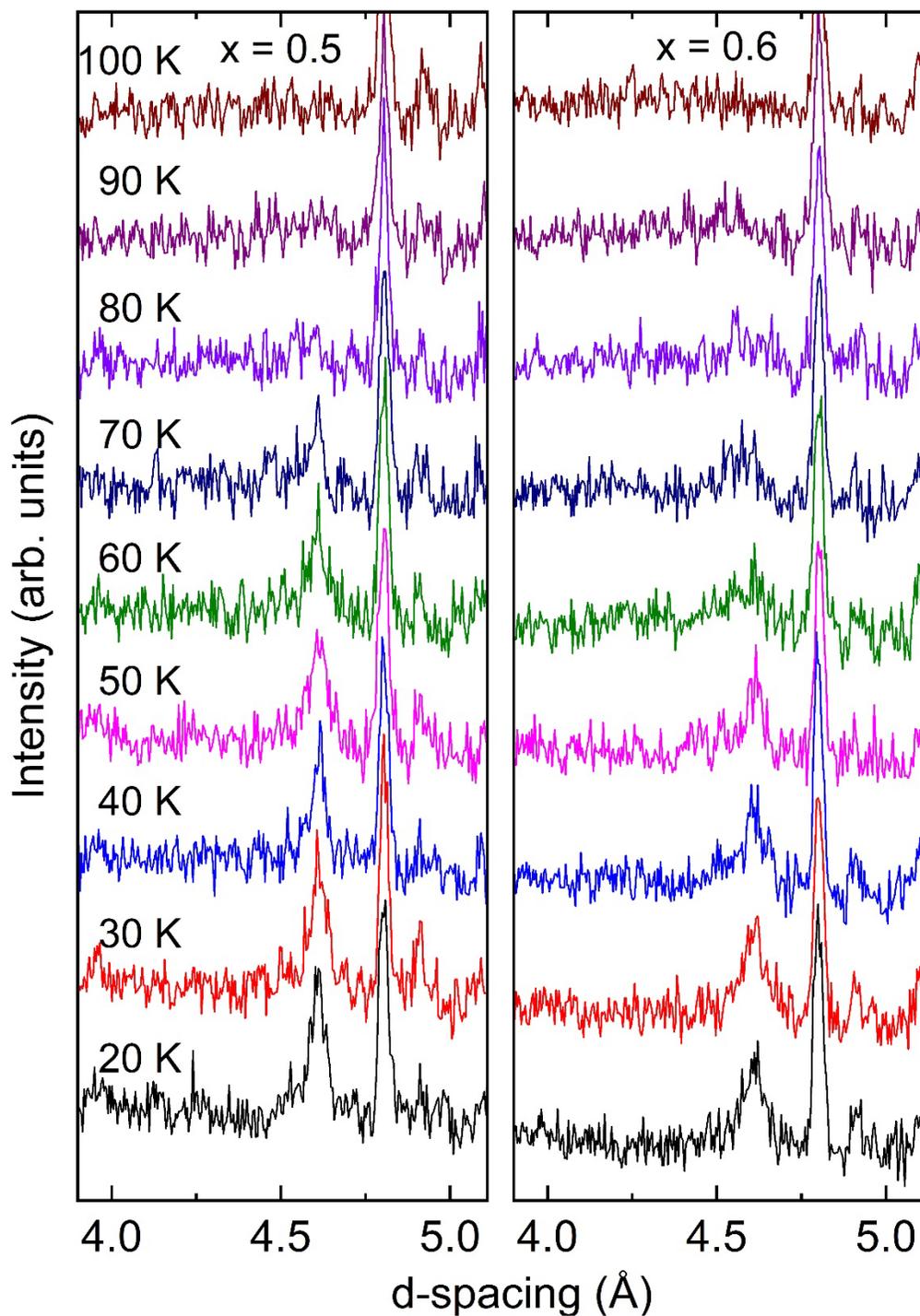

**Supplementary Figure 8.** Segments of neutron diffraction data revealing the presence of magnetic diffuse scattering for the x = 0.5 and x = 0.6 samples. The data were collected by rapidly cooling to 12 K at approximately 3.5 K/min, followed by a warming phase with data acquisition at 10 K intervals spanning from 20 K to 100 K.

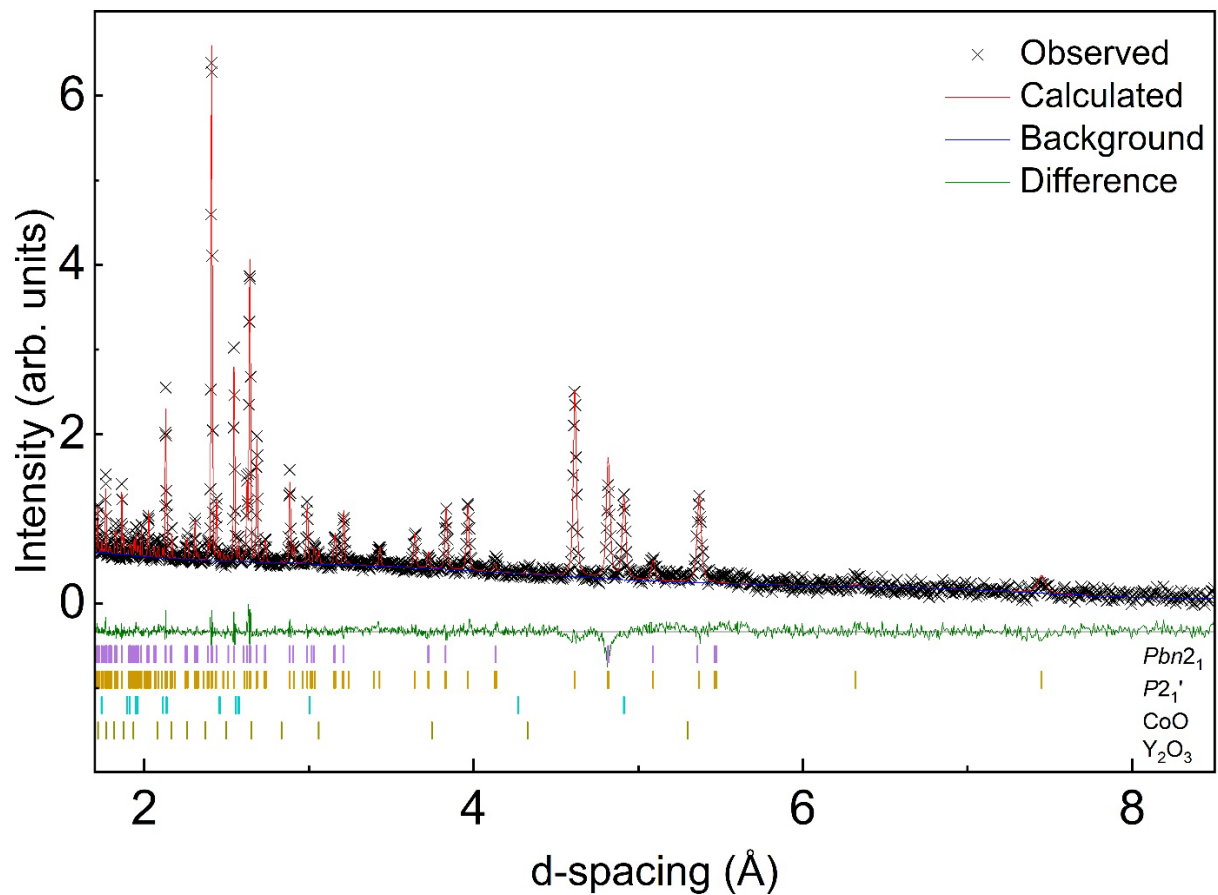

**Supplementary Figure 9**. Best-fit Rietveld refinement plot for the x = 0.1 sample at 12 K with $wR_p$ = 5.218 %. Neutron diffraction data collected after fast cooling (∼3.5 K/min) down to 12 K. Minute amounts of $Y_2O_3$ and CoO impurities were observed and included in the refinements.

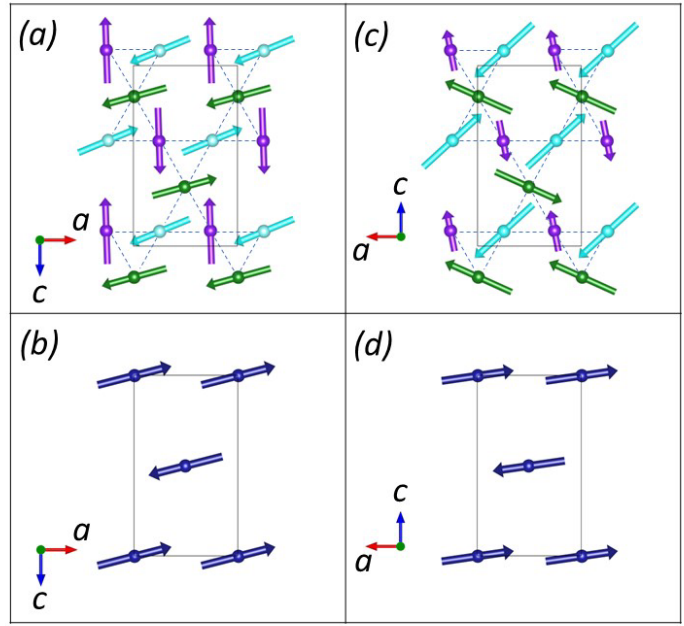

**Supplementary Figure 10.** Magnetic structures of the Kagomé and triangular layer at 5 K of the x = 0.0 sample modeled using two approaches, both achieving similar agreement parameters. In model A (*a*) and (*b*), the weighted profile residual factor wR$_p$ is 5.246%, while in model B (*c*) and (*d*), it is 5.265%.

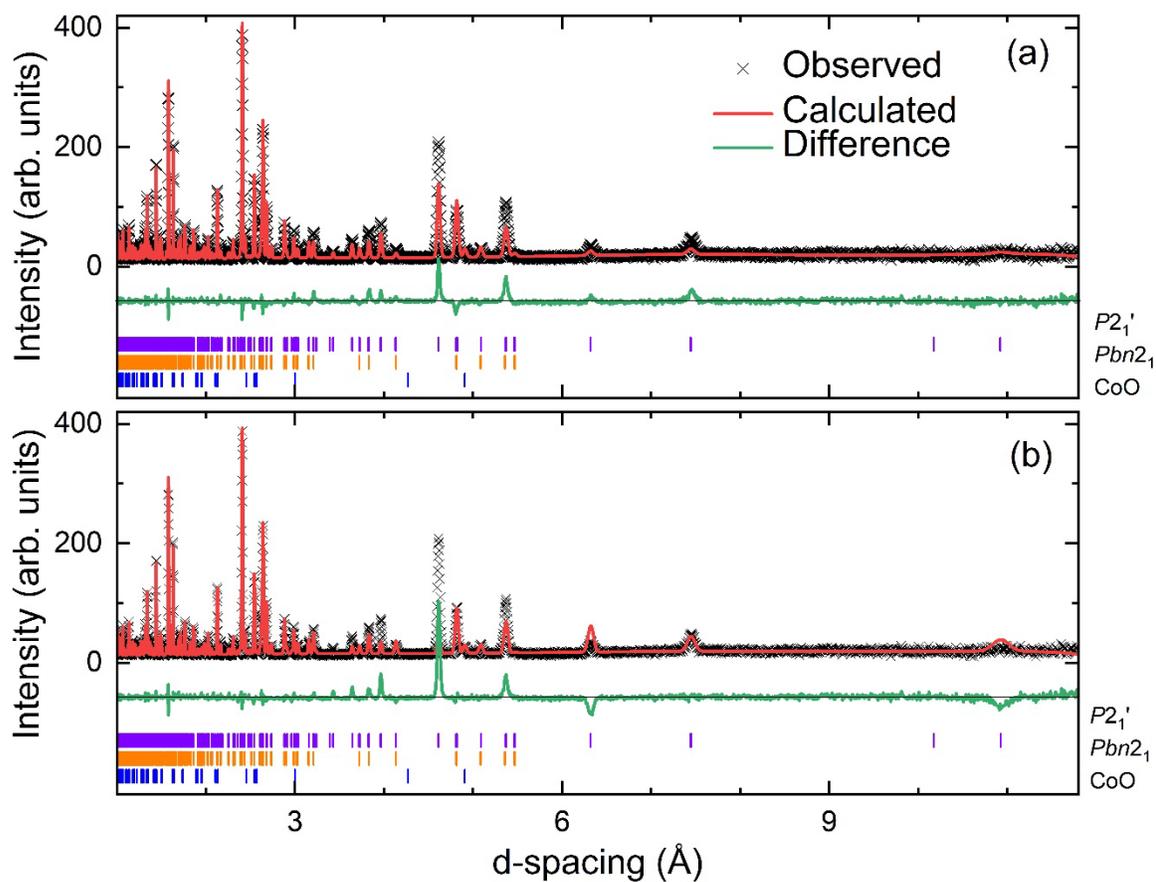

**Supplementary Figure 11**. Best-fit Rietveld refinement plots for the x = 0.0 sample at 12 K using two magnetic models: (a) Khalyavin *et al.*'s model [13] resulting in a wR$_p$ value of 6.54 %, and (b) Hoch et *al.*'s model [14], yielding a wR$_p$ value of 8.085 %.

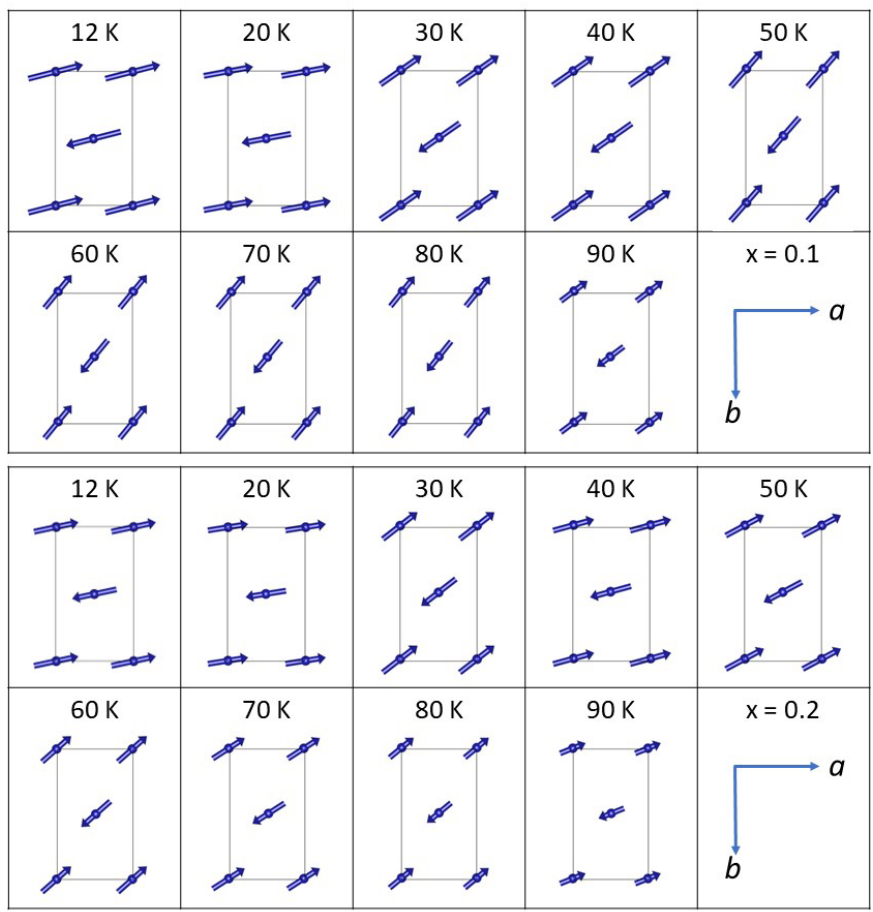

**Supplementary Figure 12** The effect of temperature on the magnetic structure of the triangular layer is displayed for the x = 0.1 and x = 0.2 samples. In agreement with the magnetic P2$_1$' space group symmetry, all the magnetic moments are confined to the ab plane. Neutron diffraction data were acquired on warming following rapid cooling at a rate of approximately 3.5 K/min down to 12 K.